\documentclass[acmsmall]{acmart}
\usepackage{subcaption}

\AtBeginDocument{%
  \providecommand\BibTeX{{%
    \normalfont B\kern-0.5em{\scshape i\kern-0.25em b}\kern-0.8em\TeX}}}

\setcopyright{acmcopyright}
\copyrightyear{2021}
\acmYear{2021}
\acmDOI{10.1145/1122445.1122456}

\acmJournal{JACM}
\acmVolume{37}
\acmNumber{4}
\acmArticle{111}
\acmMonth{8}

\begin{document}

\title{Characterizing Student Engagement Moods for Dropout Prediction in Question Pool Websites}

\author{Reza Hadi Mogavi}
\email{rhadimogavi@cse.ust.hk}
\affiliation{%
  \institution{Hong Kong University of Science and Technology}
  \country{Hong Kong SAR}}

\author{Xiaojuan Ma}
\email{mxj@cse.ust.hk}
\affiliation{%
  \institution{Hong Kong University of Science and Technology}
  \country{Hong Kong SAR}}

\author{Pan Hui}
\email{panhui@cse.ust.hk}
\affiliation{%
  \institution{Hong Kong University of Science and Technology \& University of Helsinki}
  \country{Hong Kong SAR \& Finland}}







\renewcommand{\shortauthors}{R.H. Mogavi et al.}

\begin{abstract}
Problem-Based Learning (PBL) is a popular approach to instruction that supports students to get hands-on training by solving problems. Question Pool websites (QPs) such as LeetCode, Code Chef, and Math Playground help PBL by supplying authentic, diverse, and contextualized questions to students. Nonetheless, empirical findings suggest that 40\% to 80\% of students registered in QPs drop out in less than two months. This research is the first attempt to understand and predict student dropouts from QPs via exploiting students' engagement moods. Adopting a data-driven approach, we identify five different engagement moods for QP students, which are namely \textit{challenge-seeker}, \textit{subject-seeker}, \textit{interest-seeker}, \textit{joy-seeker}, and \textit{non-seeker}. We find that students have collective preferences for answering questions in each engagement mood, and deviation from those preferences increases their probability of dropping out significantly. Last but not least, this paper contributes by introducing a new hybrid machine learning model (we call Dropout-Plus) for predicting student dropouts in QPs. The test results on a popular QP in China, with nearly 10K students, show that Dropout-Plus can exceed the rival algorithms' dropout prediction performance in terms of accuracy, F1-measure, and AUC. We wrap up our work by giving some design suggestions to QP managers and online learning professionals to reduce their student dropouts.
\end{abstract}


\maketitle

\section{Introduction}
Problem-Based Learning (PBL) is a student-centered approach to instruction where students learn through solving problems \cite{10.1145/3386527.3406751}. Question Pool websites such as LeetCode, Code Chef, Timus, Jutge, and Math Playground support PBL by supplying students with a variety of questions, quizzes, and competitions in different subjects \cite{Xia:2019:PPI:3290605.3300864,10.1007/978-3-030-23207-8_13,Wasik:2018:SOJ:3177787.3143560}. However, empirical statistics from these websites show that 40\% to 80\% of the registered students in QPs tend to drop out\footnote{Depending on the context of research, the terms \textit{attrition}, \textit{churning}, and \textit{dropping out} can be used interchangeably to imply similar concepts.} before completing their second-month of membership \cite{LeetCodesite, Code_chef, Timus_Web, Jutge_Web, MathPlay_Web}. Having said this, practical insights into this phenomenon can help educators and online learning professionals to improve their QP designs and reduce dropouts.

Nevertheless, the majority of empirical studies to date in the area of computer-mediated education are only focused on studying dropouts from Massive Open Online Courses (MOOCs) and Community Question Answering (CQA) websites \cite{srba2016comprehensive, Nagrecha:2017:MDP:3041021.3054162, mogavi2019hrcr, chen2016dropoutseer}. Therefore, there is a research gap in the literature for studying student dropouts in comparatively new platforms like QPs. Our research aims to fill this gap and inspect the problem of student dropouts in QPs through the lens of student engagement moods. By doing so, we draw Human-Computer Interaction (HCI) and Computer Supported Cooperative Work (CSCW) researchers' attention to the importance of personalization in QPs. More formally, this work answers three research questions as follows:
\begin{itemize}
    \item \textbf{RQ1:} What are student engagement moods in QPs?  
    \item \textbf{RQ2:} How are student engagement moods and dropout rates correlated?
    \item \textbf{RQ3:} Can student engagement moods help to predict student dropouts more precisely?
\end{itemize}

We utilize a probabilistic graphical model, known as Hidden Markov Model (HMM), to extract and visually distinguish different student engagement moods in QPs. We identify five dominant student engagement moods, which are \textit{(E1) challenge-seeker}, \textit{(E2) subject-seeker}, \textit{(E3) interest-seeker}, \textit{(E4) joy-seeker}, and \textit{(E5) non-seeker}. 
We distinguish each mood according to students' data-driven behavioral patterns that emerge in the process of interacting with QPs. We are inspired by the Hexad user types of Tondello et al. (see \cite{tondello2016gamification}) for naming the extracted engagement moods, but the context and concepts we introduce are genuine and specialized for QPs. 

To the best of our knowledge, this work is the first research that casts a typology for student behaviors in QPs. By adopting a data-driven approach, we identify some distinctive behavioral patterns for different QP students. For example, when students are in the challenge-seeker mood, they search for challenging types of questions that are commensurate with their high-level skills.
Students who are in the subject-seeker mood are described best as mission or task-oriented individuals. Interestingly, they do not search much to find their questions and often restrict themselves to a predefined study plan around specific subject matter and contexts. Students in this mood are more in need of a mentor, guide, or a study plan to keep them focused and help them find their questions easily. 
When students are in an interest-seeker mood, they rummage around for a variety of topics to find their questions of interest. 
However, they do not chase the challenging questions as challenge-seekers do.

Furthermore, we notice that the students in joy-seeking and non-seeking moods are not as committed as students in the other moods to study and exercise their knowledge. Joy-seeker students have a high tendency to \textit{game the platform} or misuse it for purposes other than education \cite{baker2005designing, d2008developing, baker2004off}. Technically speaking, students who exploit the platform's properties rather than their knowledge or skills to become successful in an educational platform are considered to be gaming the platform \cite{baker2007modeling}. Finally, students in a non-seeker mood tend to leave the platform earlier than students in other moods. These students are seldom determined to answer any questions. They only check the platform to see what is new and if any questions can attract their attention by chance. These findings underline the familiar point that a one-size design QP does not fit all students \cite{Teasley2017, chung2017finding, lessel2019enable}.

We also find that students have collective preferences for answering questions in each engagement mood, and deviation from those preferences increases their probability of dropping out significantly. Finally, inspired by the HMM findings and insights about student dropouts, we feed HMM results to a Long Short-Term Memory (LSTM) recurrent neural network to find if it can improve the accuracy of dropout predictions compared with a plain LSTM model and five other baselines, including an XGBoost model, Random Forest, Decision Tree (DT), Logistic Regression, and Support Vector Machine (SVM) \cite{vafeiadis2015comparison, Nagrecha:2017:MDP:3041021.3054162}. We notice improvements in the accuracy of dropout predictions when HMM and LSTM recurrent neural networks are combined. More precisely, we reach an accuracy of 78.22\%, F1-measure of 81.28\%, and AUC measure of 89.10\%, which until now set the bar for dropout prediction models in QPs. 

\textbf{Contributions.} This work is important to HCI and CSCW because it presents the first typology and dropout prediction model for QP students. Furthermore, it reinforces the need for personalizing QP websites by revealing that different students have different preferences for selecting and answering QP questions. Such knowledge could help QPs to make more informed design decisions. We also provide some design suggestions for QP managers and online learning professionals to reduce student dropouts in QPs.
\section{Background}\label{Background}
Question Pool websites (QP) provide students with a collection of questions to learn and practice their knowledge online \cite{Xia:2019:PPI:3290605.3300864}. The QPs such as Timus, Jutge.org, Optil.io, Code Chef, and HDU virtual judge are among the most popular web-based platforms for code education \cite{Petit:2012:JEP:2157136.2157267,Wasik:2018:SOJ:3177787.3143560,Xia:2019:PPI:3290605.3300864, wasik2016optil}. These platforms usually include a large repository of programming questions from which students choose to answer. The students submit their solutions to the QP and wait for the feedback to find if their code is correct. Dropout prediction is a challenging but necessary study for ensuring the sustainability and service continuation of these platforms.

\begin{figure}
     \centering
    \begin{minipage}[b]{0.48\textwidth}
         \centering
         \includegraphics[width=\textwidth]{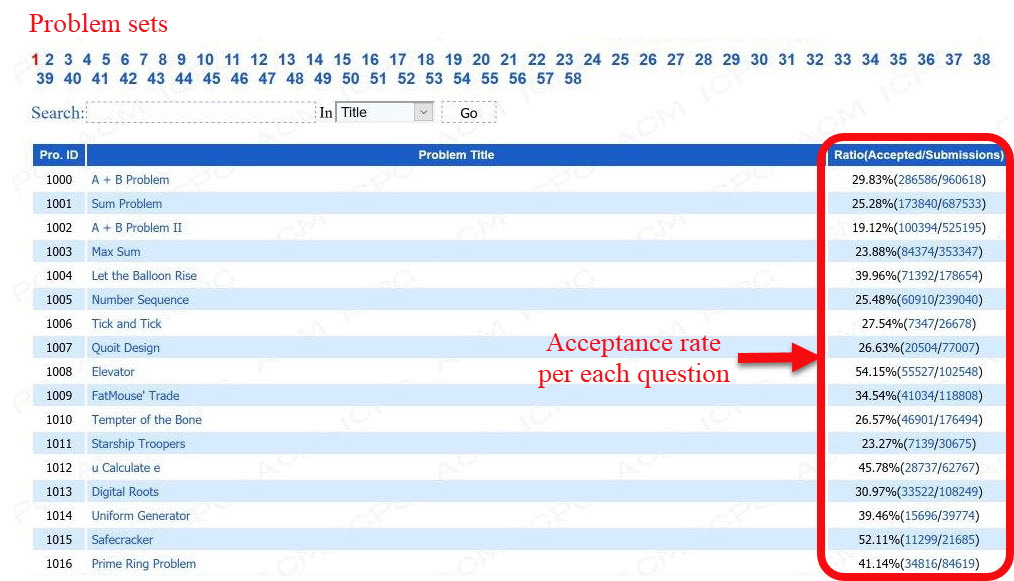}
         \subcaption{Pool of Questions}~\label{fig:HMMTCH1}
     \end{minipage}\hfill
          \begin{minipage}[b]{0.48\textwidth}
         \centering
         \includegraphics[width=\textwidth]{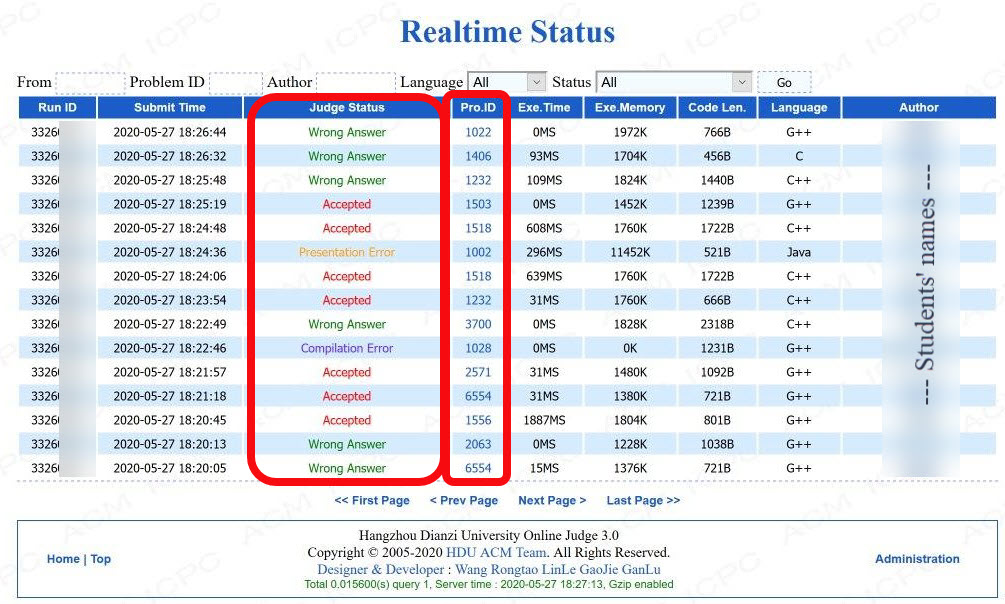}
         \subcaption{Realtime Status of the QP}~\label{fig:HMMTCU2}
     \end{minipage}
        \caption{Snapshots from HDU, the QP we study in this work. Figure (a) shows the pool of questions and each question's acceptance rate. Figure (b) shows the QP's realtime status that helps students become aware of the evaluation of their own and their friends' performances after each submission. }~\label{fig:SnapshotsHDU}
\end{figure}

In this paper, we concentrate on a popular and publicly accessible QP in China that is known as HDU virtual judge\footnote{\url{http://code.hdu.edu.cn/}} platform (henceforth HDU). The website originally belongs to Hangzhou Dianzi University's ACM team and is designed to provide students hands-on exercises to hone their programming and coding skills \cite{Xia:2019:PPI:3290605.3300864}. HDU, on average, hosts more than one hundred students every day and receives more than 300K programming code submissions every month. It is also a familiar platform for researchers who work in the field of HCI \cite{Xia:2019:PPI:3290605.3300864}. Figure \ref{fig:SnapshotsHDU} shows snapshots of HDU website.

Similar to the literature works \cite{Dror:2012:CPN:2187980.2188207,rowe2013mining,yang2018stay}, we formulate the dropout prediction problem as a binary classification task. Our definition of dropout is similar to \cite{mogavi2019hrcr}, which temporally splits a dataset into \textit{observation} and \textit{inspection} periods. The students whose number of solution submissions in the inspection period drops to less than 20\% of the observation period are considered to be dropped students. Since QP platforms usually do not have fixed start and end time points like in MOOCs \cite{Nagrecha:2017:MDP:3041021.3054162}, we regulate the observation and inspection periods similar to CQA platforms \cite{mogavi2019hrcr,Dror:2012:CPN:2187980.2188207,Pudipeddi:2014:UCF:2567948.2576965}. We use equally long periods of time for observation and inspection time windows. 
\section{Related Work}\label{RelatedWork}
\subsection{Engagement: Conceptualization and Detection Techniques}\label{R1}
The HCI and CSCW literature is abundant with various conceptualizations of engagement \cite{10.1145/3290607.3313026, foley2020student, kelders2019development, aslan2019investigating, carlton2019inferring, thebault2016exploring}. However, there is a lack of consensus on the definition of engagement \cite{barkley2009student}. The three most widely used conceptualizations of engagement moods are \textit{behavioral}, \textit{emotional}, and \textit{cognitive} \cite{fredricks2004school, alyuz2017unobtrusive}. In the context of students' learning research, behavioral engagement refers to attention, participation, and effort a student puts in doing his or her academic activities \cite{nguyen2018understanding, cappella2013classroom, alyuz2017unobtrusive}. For example, students can be in off-task or on-task moods when doing their learning tasks \cite{alyuz2017unobtrusive}. The emotional engagement refers to students' affective responses to learning activities and the individuals involved in those activities \cite{nguyen2018understanding, park2012makes}. For instance, students might be concerned about how their instructors perceive their performances. Finally, cognitive engagement is about how intrinsically invested and motivated students are in their learning process \cite{nguyen2018understanding}. For example, students might make mental efforts to debate in an online forum \cite{hind2017applying}. 

The complexity of discovering student engagement moods has resulted in the appearance of a diversity of data mining techniques \cite{bosch2016detecting, aslan2019investigating, waldrop2019measuring, moubayed2020student}. K-means, the clustering algorithm, is one of the most popular techniques many researchers use to extract naturally occurring typology of students' engagement moods \cite{moubayed2020student, saenz2011community, hamid2018dyslexia}. Saenz et al. use K-means and exploratory cluster analysis to extract different engagement moods for college students \cite{saenz2011community}. By comparison of similarities and dissimilarities between different features, they characterize 15 different clusters. Furtado et al. use a combination of hierarchical and non-hierarchical clustering algorithms to identify the contributors' profiles in the context of CQA platforms \cite{10.1145/2441776.2441916}. They categorize CQA contributors' behavior into 10 types based on how much and how well they contribute to the platform over time. However, K-Means is more useful when features show Euclidean distances properties \cite{he2015intelligence}.  

Latent variable models like Hidden Markov Models (HMMs) are also quite popular. Faucon et al. use a semi-Markov Model for simulating and capturing the behavioral engagement of MOOC students \cite{faucon2016semi}. They provide a graphical representation of the dynamics of the transitions between different states, such as forum participation or video watching. Mogavi et al. combine HMM and a Reinforcement Learning Model (RLM) to capture users' \textit{flow} experience on a famous CQA platform \cite{mogavi2019hrcr}. Flow experience is a positive mental state in psychology that occurs when the challenge level of an activity (e.g., answering a question, doing a task, or solving a homework problem) is commensurate with the user's skill level \cite{mogavi2019hrcr, pace2004grounded, csikszentmihalyi1975beyond}. Our paper is the first to use HMM to characterize students' engagement moods in QPs. Understanding students' engagement moods can help educators to manage students behaviors better and set more customized curricula \cite{shernoff2009cultivating}. 

\subsection{Dropouts: Prediction and Control}
Educational Data Mining (EDM) is a relatively new discipline that has recently caught the attention of HCI and CSCW communities \cite{costa2017evaluating, juhavnak2019using, maldonado2018mining}.
One of EDM's primary interests is to know whether students would drop out soon or continue their studies until the end of their courses or at least for a long time \cite{bassen2020reinforcement, kim2014understanding, romero2017educational}. Qiu et al. introduce a latent variable model called LadFG based on students' demographics, forum activities, and learning behaviors to predict students' success in completing the courses they start in XuetangX, one of the largest MOOCs in China \cite{qiu2016modeling}. They show that having friends on MOOCs can increase students' chances of success in receiving the final certificates of any course dramatically by three-fold, but surprisingly, being more active on the program does not guarantee the student will take the final certificate. Wang et al. propose a hybrid deep learning-based dropout prediction model that combines two architectures of Convolutional Neural Network (CNN) and Recurrent Neural Network (RNN) to advance the accuracy of dropout prediction models in MOOCs \cite{wang2017deep}. They show that their model can achieve a high accuracy comparable to feature-engineered data mining methods. As another example, Xing et al. propose a simple deep learning-based model with features such as students' access times to the platform and their history of active days to predict dropouts \cite{xing2019dropout}. They suggest finding students' dropout probability on a weekly basis to take better measures in preventing students' dropouts. Student engagement is one key factor among all of these studies. In fact, engagement can be considered a basis for students' retention, and a lack of it confronts and cancels any positive learning outcomes \cite{zainuddin2020impact}. Therefore, we use this rationale to utilize students' engagement moods to predict dropouts in QPs.
\section{Data Description}\label{DataDescription}
After receiving the approval of our local university's Institutional Review Board (IRB), we follow the ethical guidelines of AoIR\footnote{\url{https://aoir.org/ethics/}} for the study of student behavior on Hangzhou Dianzi University's QP platform (known as HDU). The dataset we study includes near 10K student records in the range from January 25th to July 15th, 2019 (172 days). We utilize the student data before April 21st for the observation period feed of the prediction models, and the data after that for the inspection of the dropouts (similar to \cite{Pudipeddi:2014:UCF:2567948.2576965,naito2018predictive}). More than half of the students drop out of the platform in the inspection period. We exclude the students with no submissions in the observation period from our study to avoid the inclusion of the students who have already dropped out and to alleviate the problem of imbalanced class labels between dropped and continuing students (see \cite{mogavi2019hrcr,Kwon:2019:GUE:3361560.3351250}). Table \ref{tab:datasummary} summarizes the main statistics of our dataset. 
\begin{table}[t!]
\small
  \centering
  \caption{The summary of the benchmark dataset}
    \begin{tabular}{p{10.835em}ccp{1em}cccc}
    \toprule
    \textbf{Measure} & \multicolumn{1}{c}{\textbf{Number}} & \multicolumn{2}{c}{\textbf{Min}} & \multicolumn{1}{c}{\textbf{Max}} & \multicolumn{1}{c}{\textbf{Median}} & \multicolumn{1}{c}{\textbf{Mean}} & \multicolumn{1}{c}{\textbf{SD}} \\
    \midrule
    Answer Submission & \multicolumn{1}{c}{1.2 M} & \multicolumn{2}{c}{1} & \multicolumn{1}{c}{62} & \multicolumn{1}{c}{5.23} & \multicolumn{1}{c}{7.14} & \multicolumn{1}{c}{18.53} \\
    Accepted Answers & \multicolumn{1}{c}{227 K} & \multicolumn{2}{c}{0} & \multicolumn{1}{c}{39} & \multicolumn{1}{c}{4.16} & \multicolumn{1}{c}{5.01} & \multicolumn{1}{c}{12.89} \\
    Endurance (in minute) & \multicolumn{1}{c}{N/A} & \multicolumn{2}{c}{0} & \multicolumn{1}{c}{209.65} & \multicolumn{1}{c}{15.91} & \multicolumn{1}{c}{21.38} & \multicolumn{1}{c}{67.43} \\
    Attendance Gap (in hour) & \multicolumn{1}{c}{N/A} & \multicolumn{2}{c}{1.18} & \multicolumn{1}{c}{673.38} & \multicolumn{1}{c}{134.22} & \multicolumn{1}{c}{192.48} & \multicolumn{1}{c}{125.68} \\
    \midrule
    \multicolumn{3}{p{17.615em}}{Number of Students = 9,941 } & \multicolumn{5}{p{14.775em}}{Number of Dropped Students = 5,261} \\
    \bottomrule
    \end{tabular}%
  \label{tab:datasummary}%
\end{table}%
\section{Research Overview}\label{Methodology}
We use an unsupervised HMM for decoding student engagement moods, and a supervised LSTM network for predicting dropouts. Both components are trained with the student data features during the observation period. The HMM inputs include simple observable features that imply student \textit{performance}, \textit{challenge}, \textit{endurance}, and \textit{attendance gap} states after each submission to the QP. The parameters of the HMM are estimated by iterations of a standard Expectation Maximization (EM) algorithm known as Baum-Welch \cite{baum1970maximization}.  Five dominant student engagement moods are identified, which are (E1) challenge-seeker, (E2) subject-seeker, (E3) interest-seeker, (E4) joy-seeker, and (E5) non-seeker. We run a user study with 26 local students to evaluate our HMM findings. 

After decoding the student engagement moods with the HMM, we associate the platform questions with the engagement moods that are more likely to submit solutions for those questions. We notice that students have collective preferences for answering questions in each engagement mood, and deviation from those preferences increases their probability of dropping out significantly. 

Finally, we feed the generated engagement moods and questions' associativity features along with other common features to an LSTM network to predict student dropout in the inspection period. All the dropout predictions are reported based on the 10-fold cross-validation. 

\section{Students' Engagement Moods (RQ1)}
\subsection{Model Construction}
Hidden Markov Models (HMMs) are statistical tools that help to make inferences about the latent (unobservable) variables through analyzing the manifest (observable) features \cite{Kokkodis:2019:RDT:3308558.3313479,shi2019state,chen2018hidden}. In the context of education systems, HMMs are used for the detection of various phenomenons such as \textit{student social loafing} \cite{Zhang:2017:ISS:3041021.3054145}, \textit{flow zone engagement} \cite{mogavi2019hrcr}, and \textit{academic pathway choices} \cite{Stanford_latent_variable_Models}. The widespread use of HMMs in the literature and the capability of capturing complex data structures \cite{Stanford_latent_variable_Models} inspire our work to apply HMMs to help distinguish between different student engagement moods in the QP platforms. We use the \textit{hmmlearn} module in Python to train our model.

\noindent$\bullet$ \textbf{Inputs.} In order to build an HMM, we utilize the manifest features for student \textit{performance}, \textit{challenge}, \textit{endurance}, and \textit{attendance gap}. We pick the manifest features by performing an extensive thematic analysis \cite{Parsons:2015:TAS:2740908.2741721,ward2009developing} across the literature about student engagement \cite{hamari2016challenging,Hoffman:2019:SEK:3287324.3287438,Morgan:2018:CAV:3197091.3197092}. The features we use are as follows.
\begin{itemize}
    \item[-]\underline{Performance}: The feedback QP platforms provide each submission, such as if the answer is \texttt{Wrong} or \texttt{Accepted}.
    \item[-]\underline{Challenge}: The past acceptance rate of a question, which is often shown along with a guide next to each question, can resemble the challenge. 
    \item[-]\underline{Endurance}: The time students spend on the platform to answer questions and compile codes in one session is student endurance. Similar to \cite{Hara:2018:DAW:3173574.3174023}, we define a ``session'' in QP platforms as a period of time in which the interval between two consecutive submissions does not exceed one hour. We measure the student endurance in minutes.
    \item[-]\underline{Attendance gap}: The time interval between two consecutive sessions is a student's attendance gap. We measure the attendance gap in hours.
\end{itemize}
We should mention here that these features serve only as cues to infer students' cognitive (performance and challenge) and behavioral (endurance and attendance gap) engagement moods \cite{kuh2001assessing, handelsman2005measure, ouimet2005assessment, langley2006student}.

\noindent$\bullet$ \textbf{Parameters.} Hereafter, we use a four-element vector \scalebox{0.90}{$O_{t}$} to refer to the corresponding student observations after each answer submission to the QP in time \scalebox{0.90}{$t$}. 
The HMM assumes the observations \scalebox{0.90}{$O_{t}$} are generated by an underlying state space of hidden variables \scalebox{0.90}{$Z=\{z_{i}\}$}, where \scalebox{0.90}{$i\geq 1$}.
For convenience, we use a triplet \scalebox{0.90}{$\lambda_{\scalebox{0.50}{HMM}}=(A, B, \pi)$} to denote the HMM we train for extraction of the student engagement moods. The transition matrix \scalebox{0.90}{$A$} shows the probabilities of moving between different engagement moods over time. The emission matrix \scalebox{0.90}{$B$} shows the conditional probability for an observation \scalebox{0.90}{$O_{t}$} to be emitted (generated) from a certain engagement mood \scalebox{0.90}{$z$}. The vector \scalebox{0.90}{$\pi$} denotes the initial probabilities of being in each of the engagement moods of \scalebox{0.90}{$Z$}. The initial probabilities are often assumed to be \scalebox{0.90}{$\frac{1}{||Z||}$}, with \scalebox{0.90}{$||Z||$} showing the cardinality of the hidden state space (i.e., the number of engagement moods) \cite{mogavi2019hrcr}.

\noindent$\bullet$ \textbf{Model training.} We optimize \scalebox{0.90}{$\lambda_{\scalebox{0.50}{HMM}}$} parameters according to the student behavior in the observation period with a standard EM algorithm known as Baum-Welch \cite{baum1970maximization}. The aim is to optimize the
\scalebox{0.90}{$\lambda_{\scalebox{0.50}{HMM}}$} parameters such that \scalebox{0.90}{$Pr(\lambda_{\scalebox{0.50}{HMM}}|O)$} is maximized with \scalebox{0.90}{$O=\{O_{t}\}$}. To avoid the local maximum problem with the EM algorithm, we train the HMM with ten random seeds until they converge at the global maximum. 

\begin{figure}
     \centering
         \includegraphics[width=0.45\textwidth]{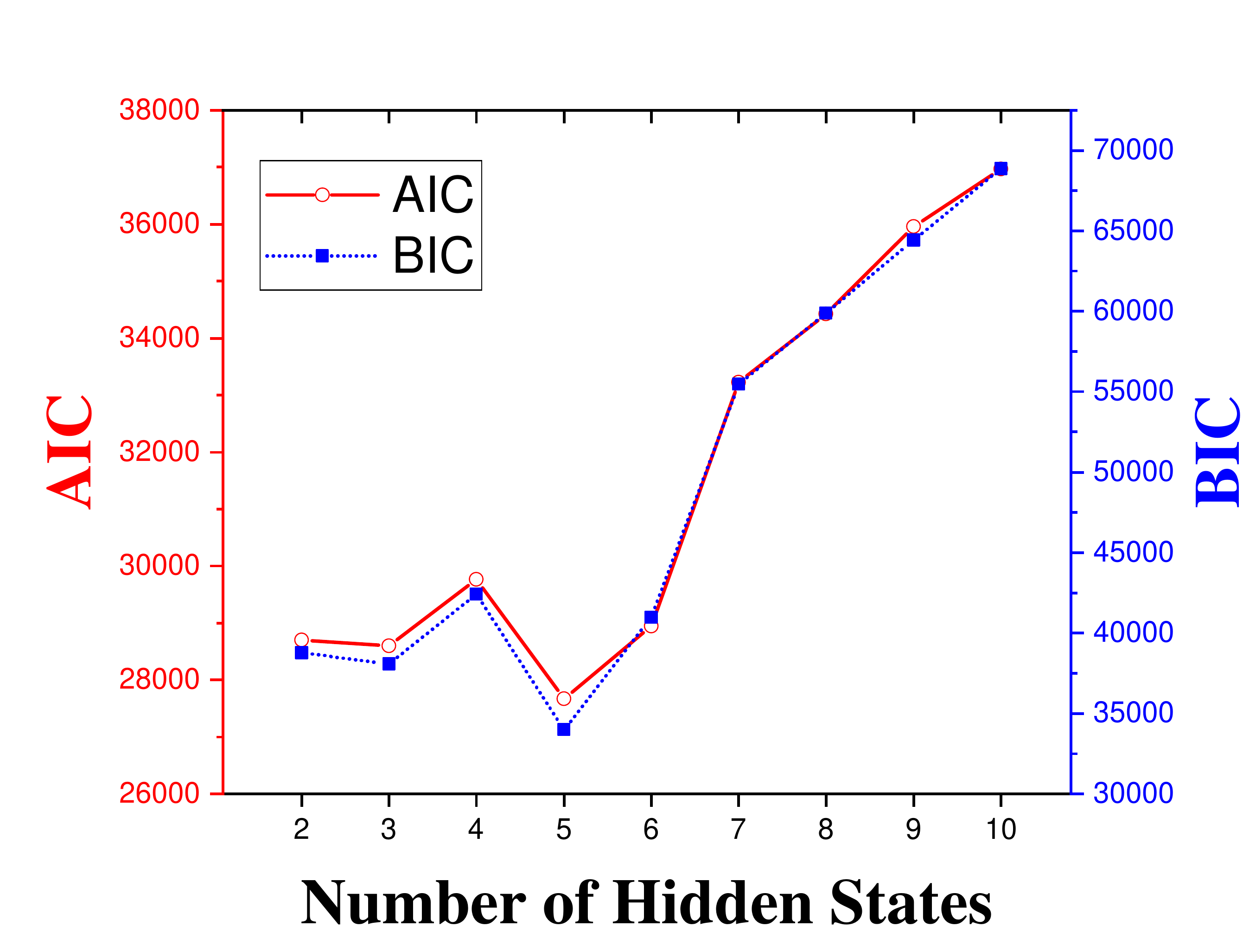}
         \caption{Estimating the best number of hidden states, where AIC and BIC measures both take the least values.}~\label{fig:HMMestimation}
\end{figure}

The HMM representation is completed by choosing the best number of hidden states \cite{mogavi2019hrcr}. While this task appears to be simple conceptually, finding the best number of hidden states in a meaningful way is quite challenging \cite{van2011process,mogavi2019hrcr}. The main reason is that an HMM with a small number of hidden states cannot capture the underlying behavioral kinetics adequately, and an HMM with too many hidden states is difficult to interpret \cite{okamoto2017analyzing}. However, we need a criterion to compromise, and thus we use the conventional Akaike (AIC) \cite{SHANG20082004} and Bayes (BIC) \cite{chen1998speaker} measures in our model training (also see \cite{mogavi2019hrcr}). Figure \ref{fig:HMMestimation} plots the AIC and BIC measures against the number of hidden states in our trained HMMs (i.e., from 2 to 10 hidden states are tested). We choose an HMM with \scalebox{0.90}{$||Z||=5$} hidden states since the global values of AICs and BICs measures are the lowest in this representation. The smaller AIC and BIC show more descriptive and less complicated \scalebox{0.90}{$\lambda_{\scalebox{0.50}{HMM}}$} \cite{mogavi2019hrcr}. 

\subsection{Clustering Analysis: Pattern Discovery}
\begin{figure}[t!]
     \centering
         \begin{subfigure}[b]{0.33\textwidth}
         \centering
         \includegraphics[width=\textwidth]{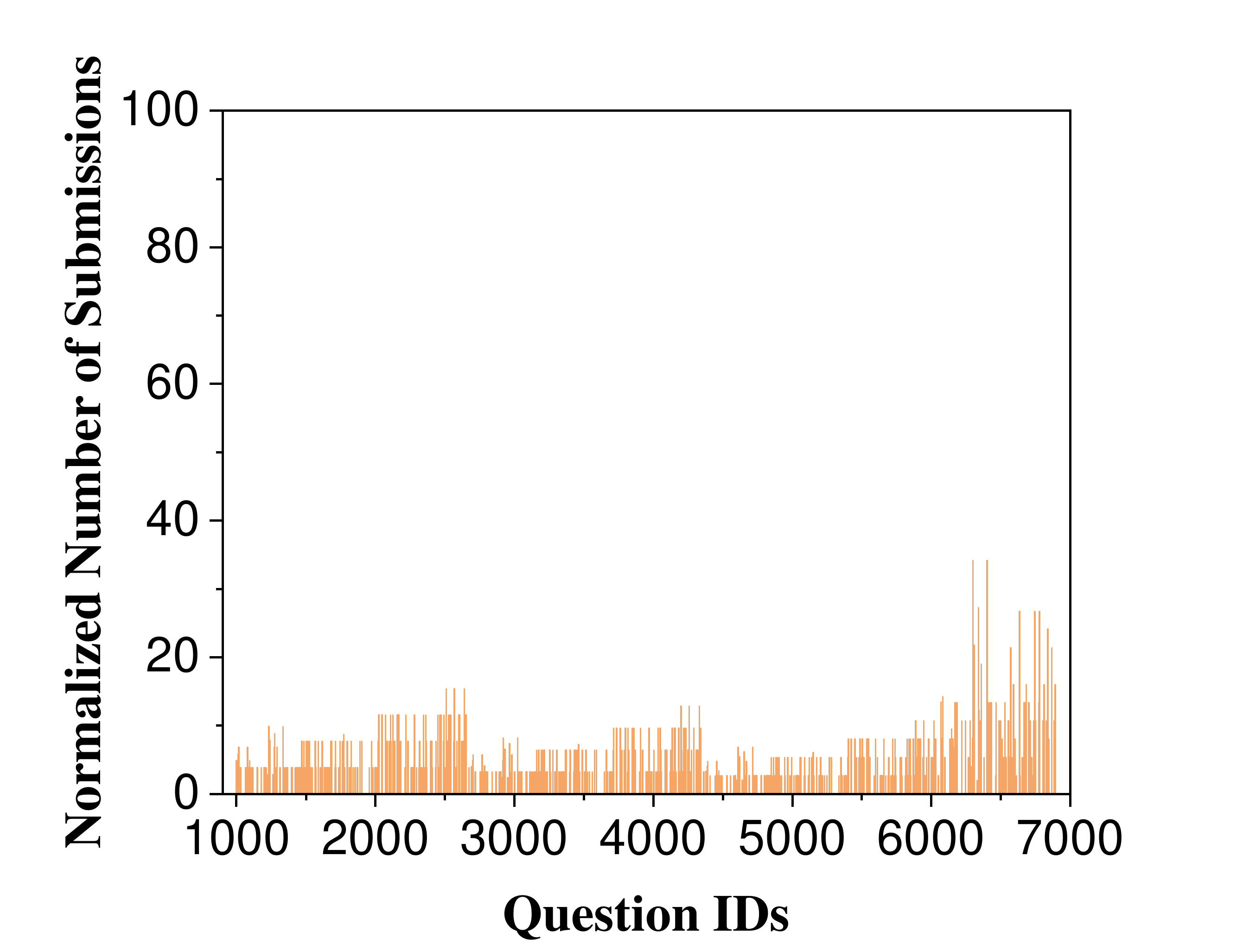}
         \caption{\scalebox{0.90}{Challenge-seeker}}~\label{fig: challenge-seeker1}
     \end{subfigure}\hfill
          \begin{subfigure}[b]{0.33\textwidth}
         \centering
         \includegraphics[width=\textwidth]{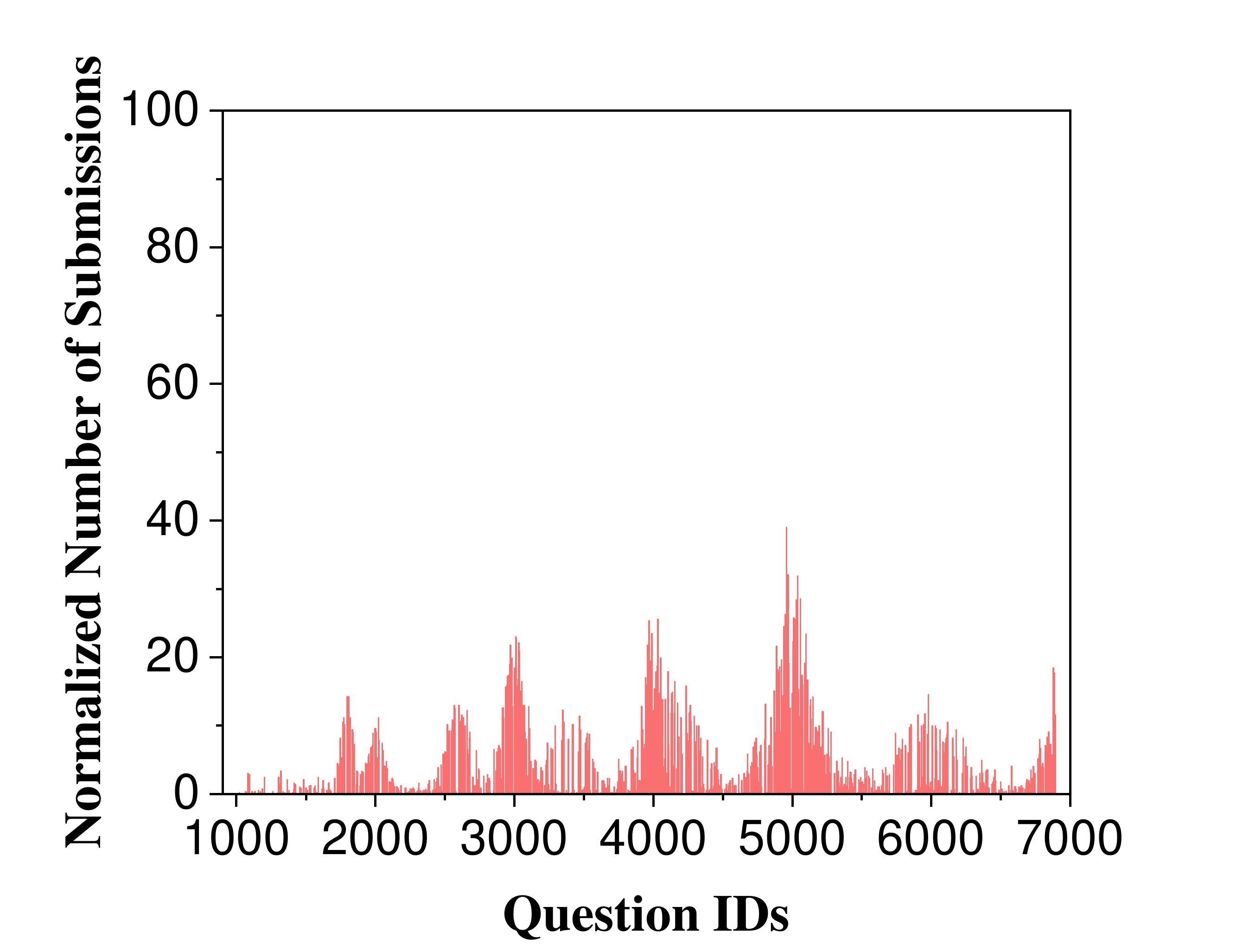}
         \caption{\scalebox{0.90}{Subject-seeker}}~\label{fig: Subject-seeker}
     \end{subfigure}\hfill
          \begin{subfigure}[b]{0.33\textwidth}
         \centering
         \includegraphics[width=\textwidth]{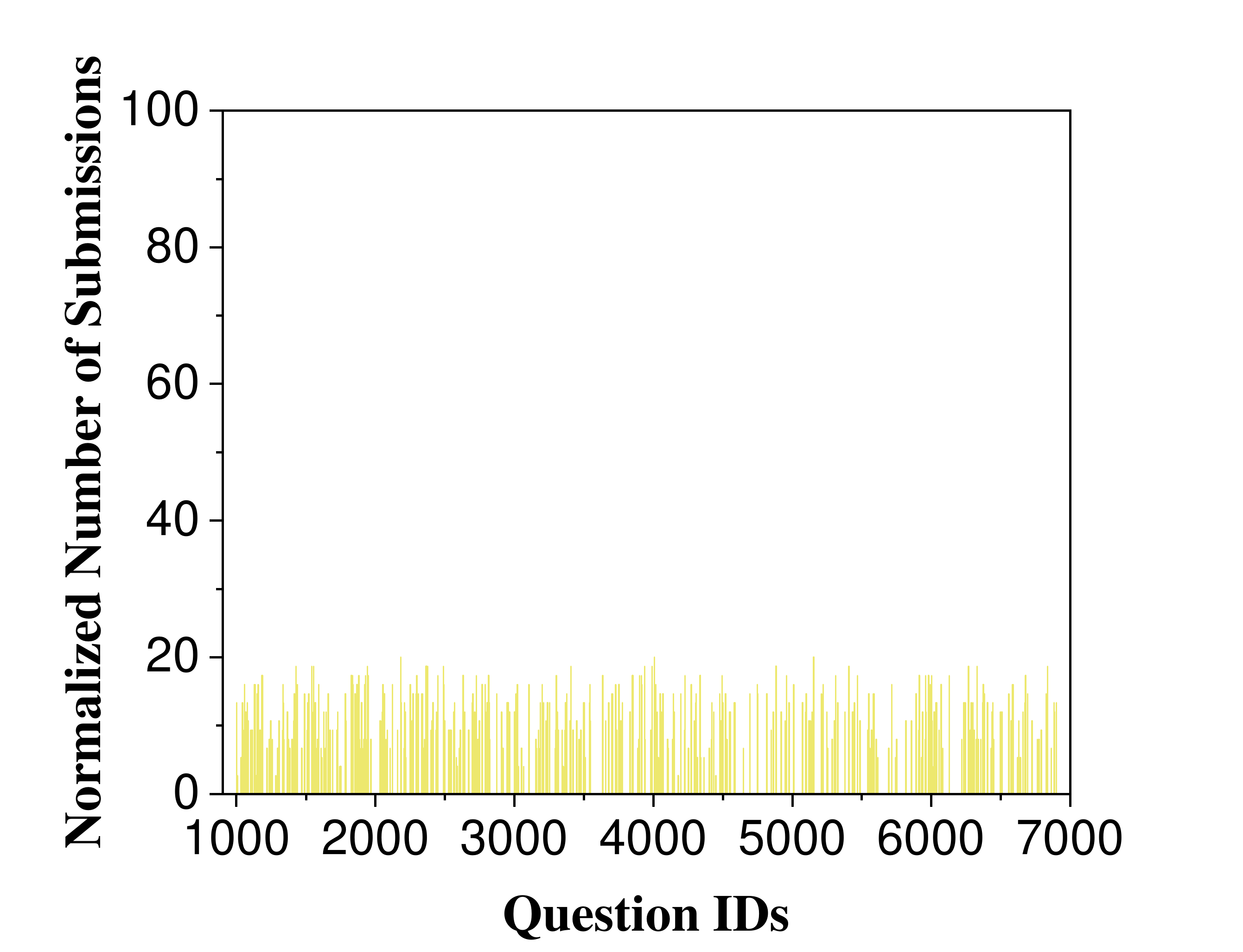}
         \caption{\scalebox{0.90}{Interest-seeker}}~\label{fig: Interest-seeker}
     \end{subfigure}\\
     \begin{subfigure}[b]{0.33\textwidth}
         \centering
         \includegraphics[width=\textwidth]{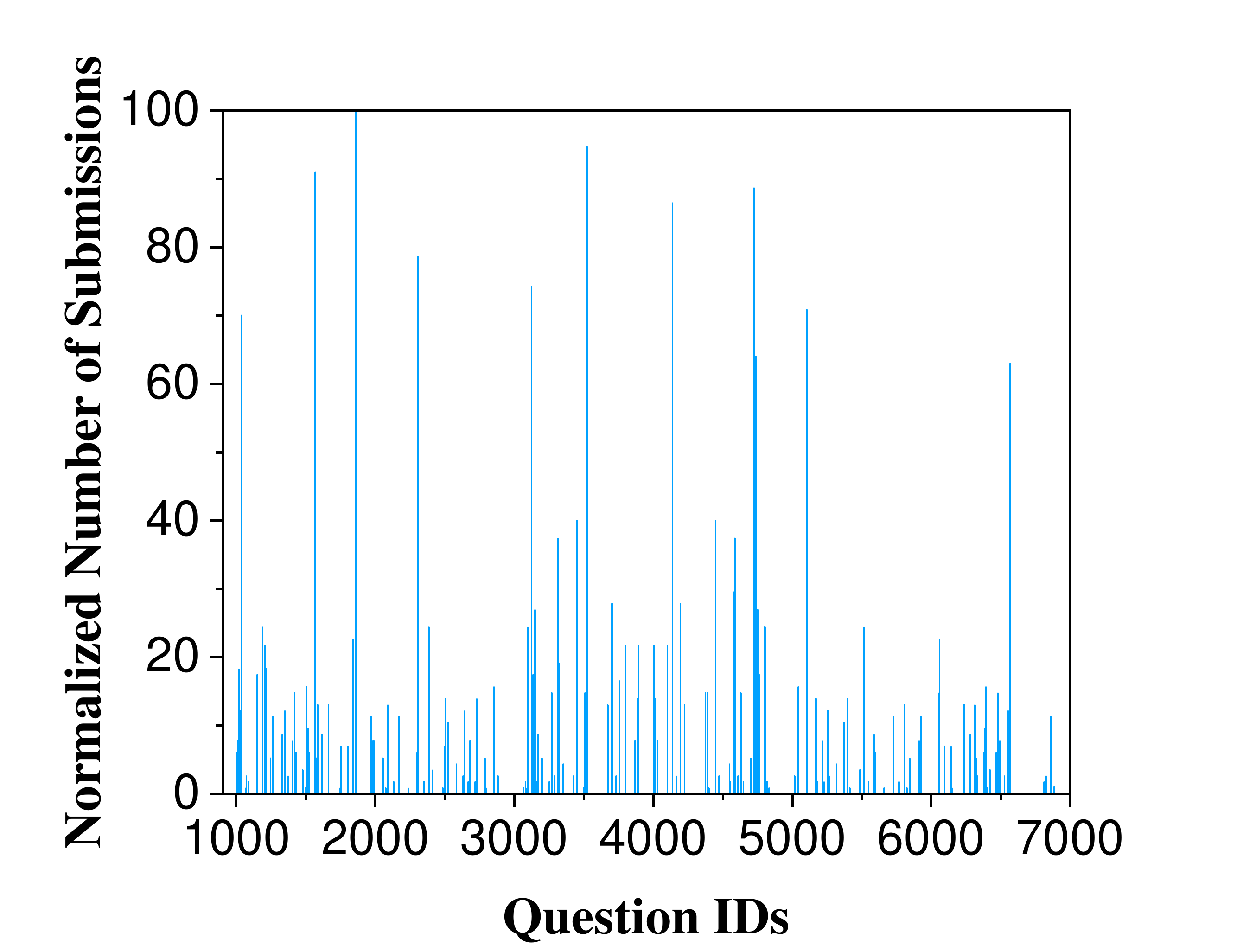}
         \caption{\scalebox{0.90}{Joy-seeker}}~\label{fig: Joy-seeker}
     \end{subfigure}
          \begin{subfigure}[b]{0.33\textwidth}
         \centering
         \includegraphics[width=\textwidth]{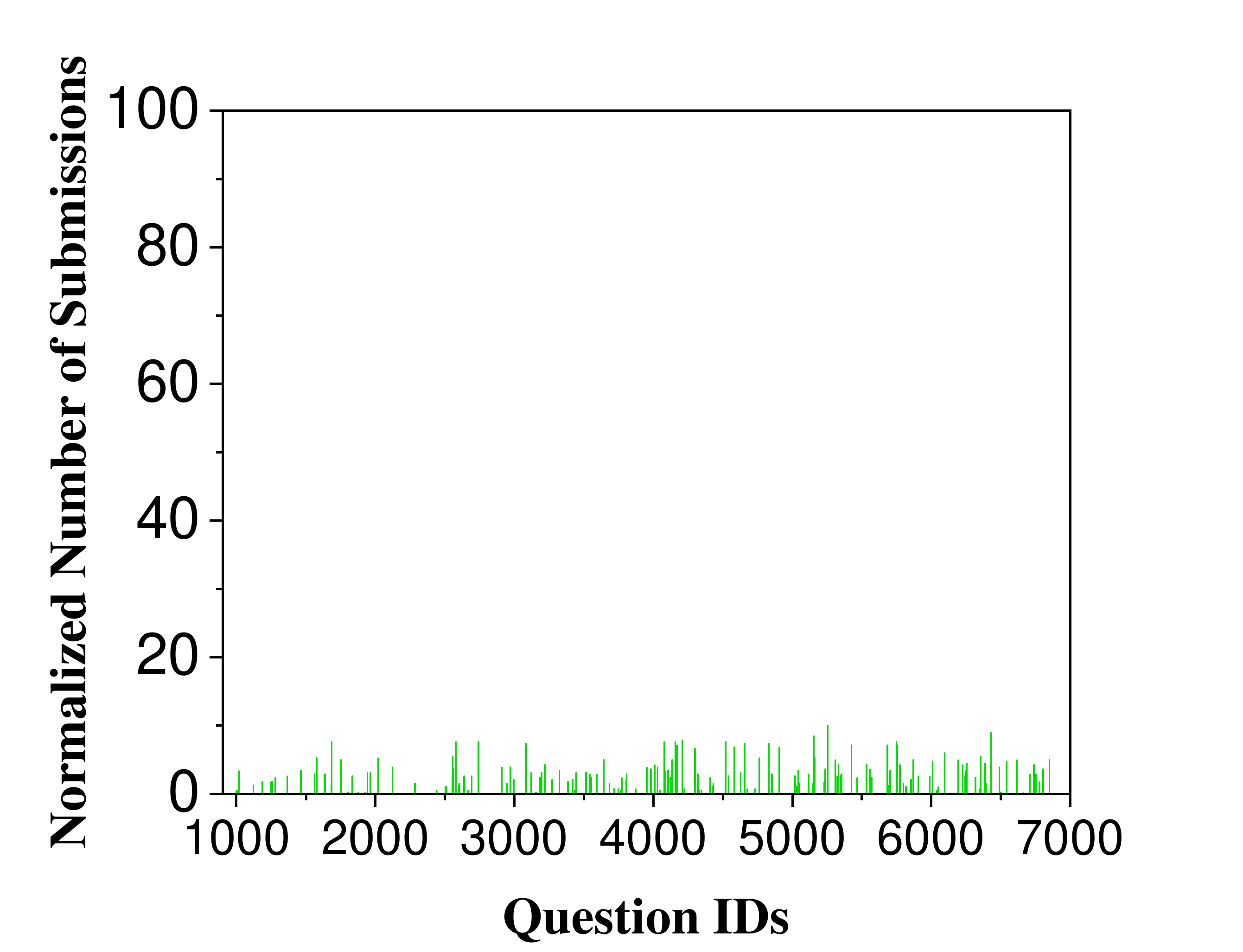}
         \caption{\scalebox{0.90}{Non-seeker}}~\label{fig: non-seeker1}
     \end{subfigure}\hfill
~\caption{Frequency plots of student submissions over different question IDs.}\label{fig: histha}
\end{figure}

Similar to the literature, we use data distributions within each hidden state to characterize and visualize the distinction between different engagement moods \cite{10.1145/2441776.2441916, mogavi2019hrcr, babbie2016basics}. The features we inspect here are the number of incorrect and accepted answers, the average ease of questions, the average time spent in the platform, the average time gap in attendance, and the number of repeated submissions. They are inspired by the manifest features we had before, but instead of an individual student, they aim to examine all students' collective behaviors in a specific hidden state. The cumulative distribution function (CDF) of these features are plotted in Figure \ref{salamCDF}. We also utilize the frequency plots of student submissions over different question IDs to demonstrate question-answering patterns in each hidden state (Figure \ref{fig: histha}). The number of submissions in each hidden state is a value normalized between 0 to 100. This representation is to facilitate the visualization and comparison of the patterns in different hidden states.
The marked characteristics of each hidden state are as follows:
\begin{figure}[t!]
     \centering
     \begin{subfigure}[b]{0.33\textwidth}
         \centering
         \includegraphics[width=\textwidth]{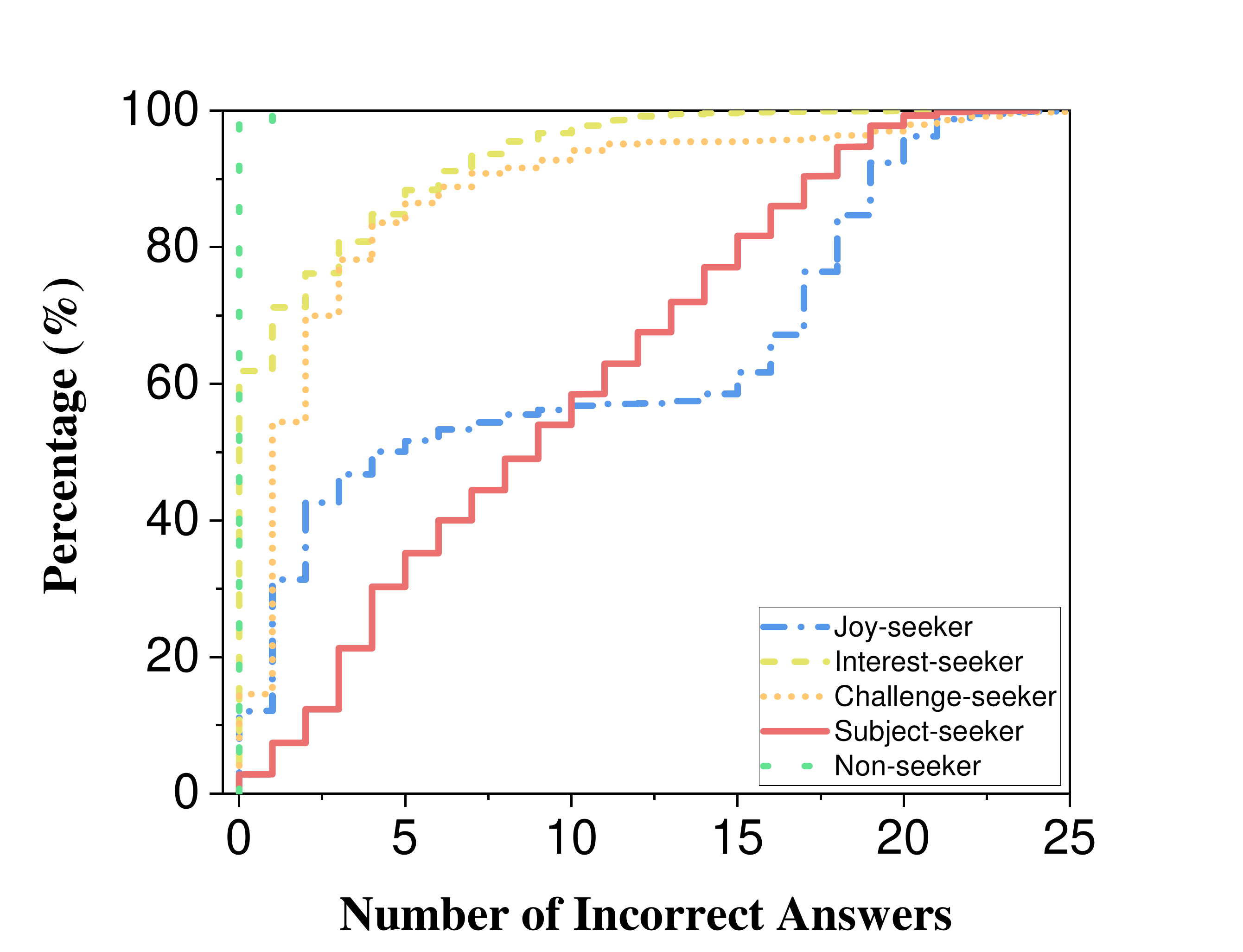}
         \caption{Incorrect Answers}~\label{fig:incans}
     \end{subfigure}\hfill
     \begin{subfigure}[b]{0.33\textwidth}
         \centering
         \includegraphics[width=\textwidth]{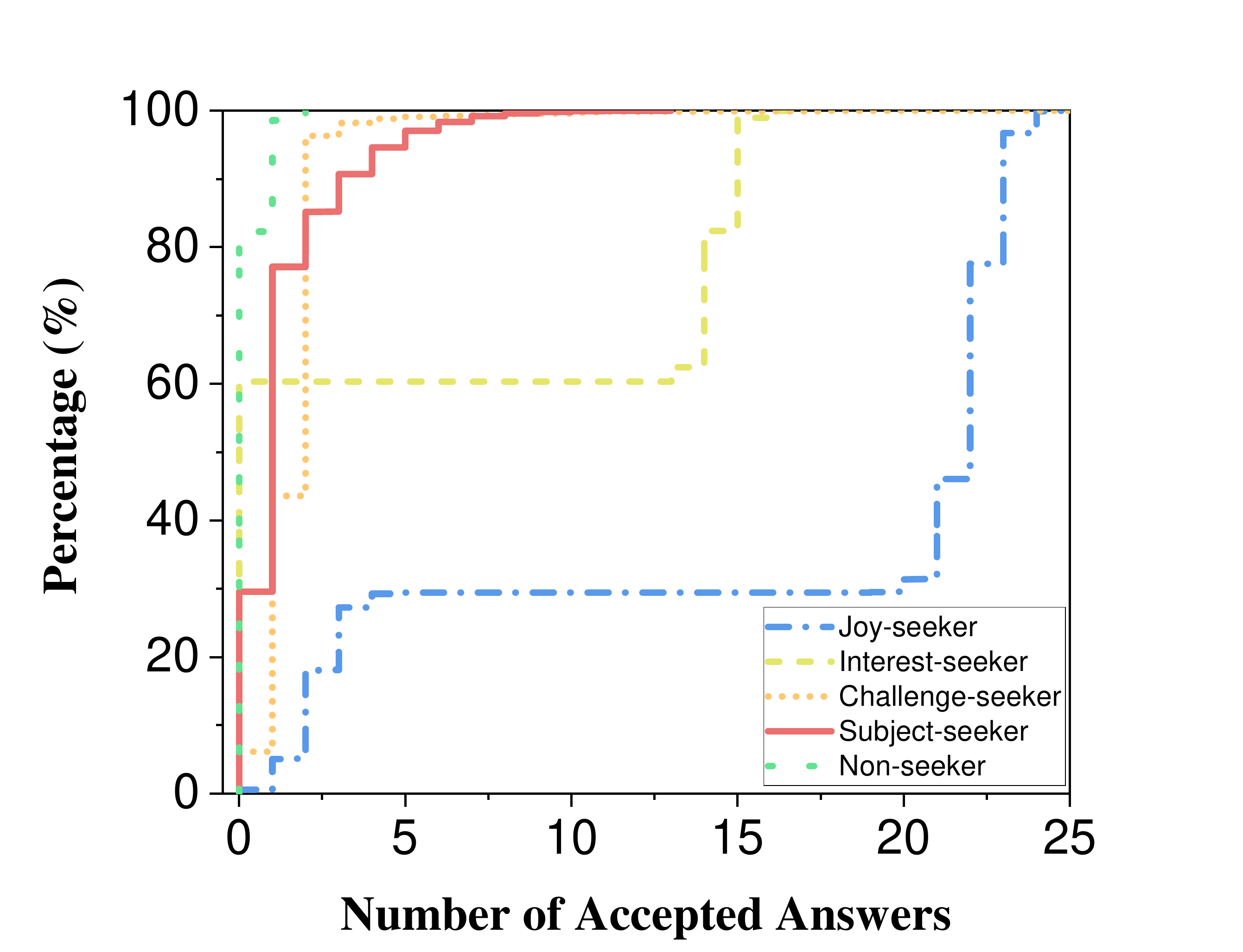}
         \caption{Accepted Answers}~\label{fig:accans}
     \end{subfigure}
     \begin{subfigure}[b]{0.33\textwidth}
         \centering
         \includegraphics[width=\textwidth]{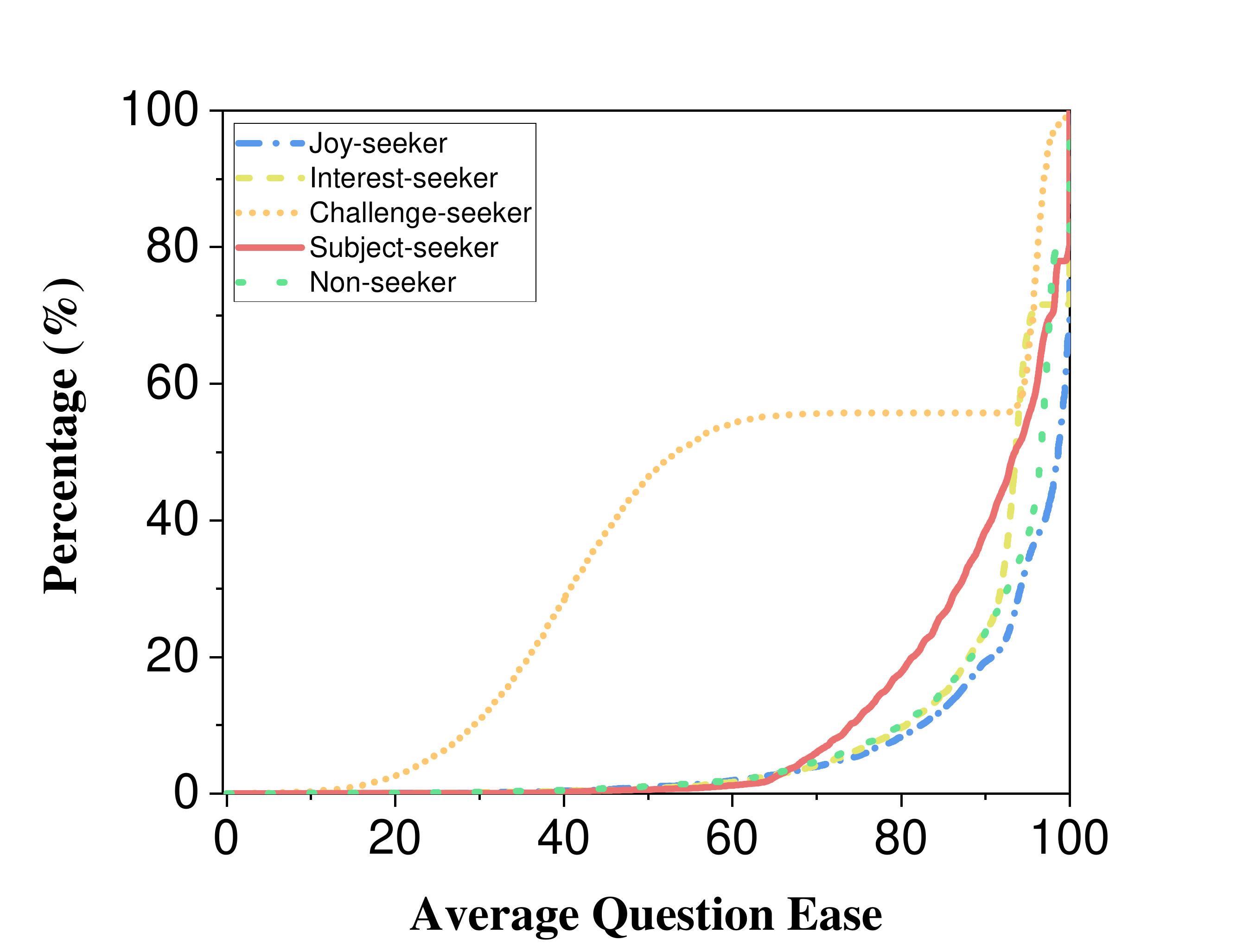}
         \caption{Question Ease}~\label{CDF_Ease}
     \end{subfigure}\hfill
     \\
         \begin{subfigure}[b]{0.33\textwidth}
         \centering
         \includegraphics[width=\textwidth]{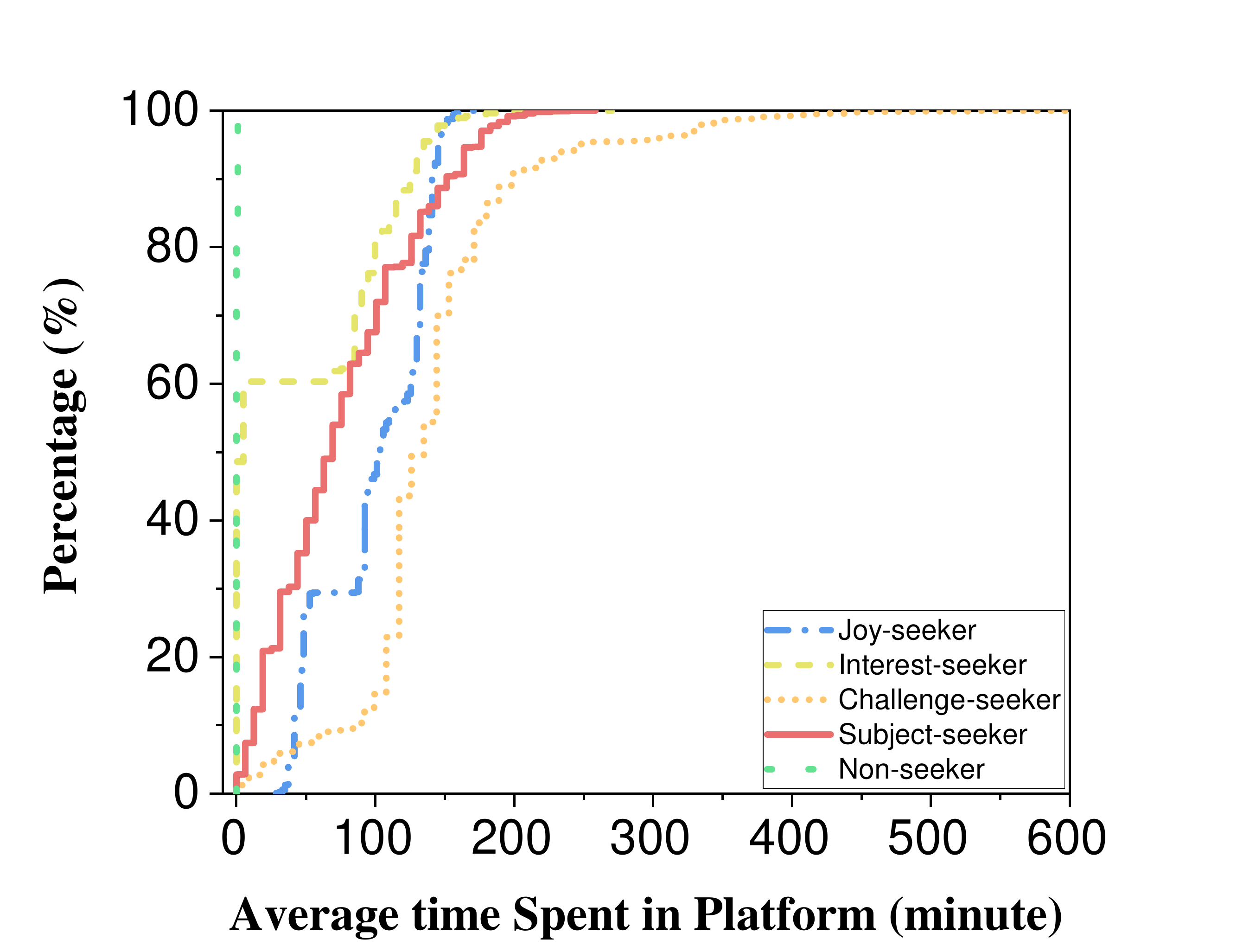}
         \caption{Spent Time}~\label{CDF_AVGTime}
     \end{subfigure}\hfill
          \begin{subfigure}[b]{0.33\textwidth}
         \centering
         \includegraphics[width=\textwidth]{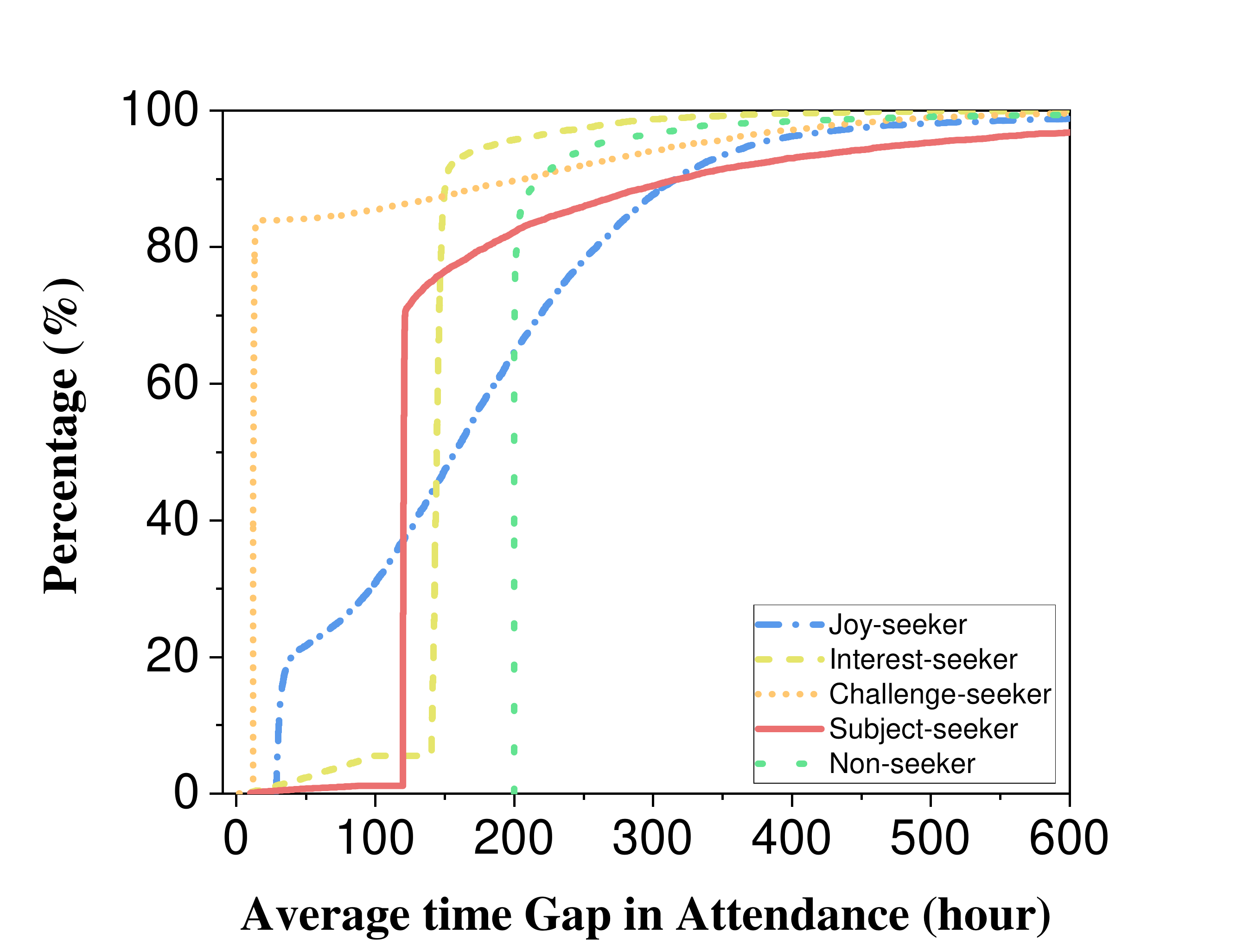}
         \caption{Gap in Attendance}~\label{CDF_GAP}
     \end{subfigure}
          \begin{subfigure}[b]{0.33\textwidth}
         \centering
         \includegraphics[width=\textwidth]{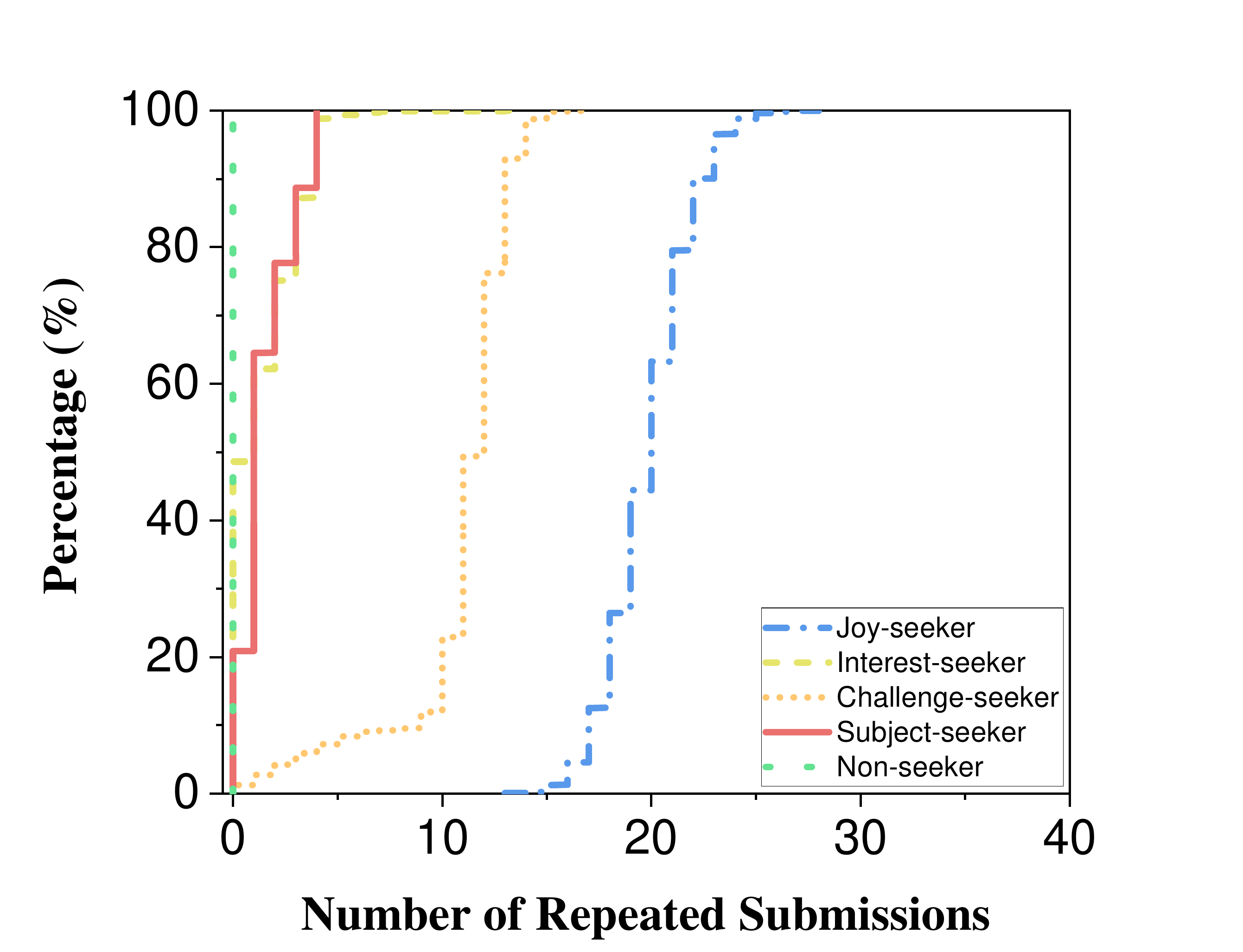}
         \caption{Repeated Submissions}~\label{fig:cdfrep}
     \end{subfigure}\hfill
~\caption{CDF plots of the student engagement moods resolved}\label{salamCDF}
\end{figure}


\noindent$\bullet$ \textbf{(E1) Challenge-seeker (Hidden State 1)}: As shown in Figure \ref{CDF_Ease}, the students in this mood are best described for their tendency to look for more challenging questions, i.e., the questions with the least acceptance rate. From Figures \ref{CDF_AVGTime} and \ref{CDF_GAP}, we find they also spend the longest average time on the platform, and attend the platform more frequently in comparison with the other engagement moods. Furthermore, Figure \ref{fig: challenge-seeker1} shows that challenge-seekers show more interest in the last questions of the platform.\\
\noindent$\bullet$ \textbf{(E2) Subject-seeker (Hidden State 2)}: As shown in Figure \ref{fig: Subject-seeker},  students in this mood tend to answer specific sets of questions. They usually answer the specific-context questions (e.g., greedy algorithms) sequentially. Furthermore, from Figure \ref{CDF_AVGTime} we can notice that subject-seekers come second after the students in the challenge-seeker mood for spending the longest time on the platform. Figures \ref{fig:incans} and \ref{fig:accans} show that the students in the subject-seeker mood have the largest average number of incorrect answers on the platform, whereas their average number of accepted answers is quite similar to those of the challenge-seeker students (the distributions are also similar).\\
\noindent$\bullet$ \textbf{(E3) Interest-seeker (Hidden State 3):} From Figure \ref{fig: Interest-seeker}, we realize that the students in this mood do not answer specific-context questions and regularly search for their questions of interest. The question types they answer has the largest variance.
Furthermore, the distribution of the easiness of the questions shown in Figure \ref{CDF_Ease} makes no tangible difference in comparison with the other moods except the challenge-seeker mood. As shown in Figure \ref{fig:accans}, the students in this mood hold the highest average number of accepted answers after the students in the joy-seeker mood. Moreover, according to Figure \ref{fig:incans}, interest-seekers' distribution of producing incorrect answers is close to a uniform distribution, which sharply distinguishes them from the other student moods.\\
\noindent$\bullet$ \textbf{(E4) Joy-seeker (Hidden State 4):} As shown in Figures \ref{CDF_Ease} and \ref{fig:cdfrep}, the students who are in this mood tend to answer the easiest questions on the platform in a highly repetitive manner. Interestingly, these students choose their questions from a small and selective number of QP questions (probably those with compilation loopholes) (see Figure \ref{fig: Joy-seeker}). Also, their number of accepted answers has the highest value among all the other moods (see Figure \ref{fig:accans}). Based on these signs, we come to the conclusion that these students are at the highest risk of \textit{gaming the platform} \cite{baker2004off, baker2005designing} in comparison with the other mood groups. Finally, we notice that the distribution of incorrect answers for this mood type resembles a mixture of two Gaussian distributions, which distinguishes itself in sharp contrast to the other mood groups (see Figure \ref{fig:incans}).

\noindent$\bullet$ \textbf{(E5) Non-seeker (Hidden State 5):} According to Figure \ref{CDF_GAP}, we notice that the students in this mood are well-distinguished by holding the largest average time gap among all moods for attending the platform, which means that they seldom visit the platform.
Furthermore, the least average time spent on the platform is also another characteristic of this mood (see Figure \ref{CDF_AVGTime}). Finally, as is expected, the lowest number of incorrect answers, accepted answers, and repeated submissions are the outcomes of this short visit (see Figures \ref{fig:incans}, \ref{fig:accans}, and \ref{fig:cdfrep}).\\ 

\subsection{HMM Evaluation: Experience Sampling}
With the approval of our university's Institutional Review Board (IRB), we conduct a preliminary user study with 26 local students to evaluate the accuracy of the hidden states resolved. A combination of snowball and convenience sampling methods is used to recruit participants from our local universities. The participants include 9 females and 17 males, all undergraduate computer-major students in the age range between 18 to 23 (mean $=$ 19.46, SD $=$ 1.36). They have taken introductory programming courses (i.e., CS1 and CS2 courses \cite{10.1145/1734263.1734335}) in the previous semester, and now use the QP website to hone their problem-solving and algorithm design skills. Thirteen students have never used any QPs before, and the others have less than one year of experience with QPs in subjects other than programming. All of the participants are asked to sign an informed consent form before we begin our study. Due to the pandemic restrictions (Covid-19), participants are asked to take part in our study from their homes' safety and comfort. As a token of our appreciation, we compensate each participants' time with small cash (50HK\$) at the end of our study.   

We use the event-focused version of Experience Sampling Method (ESM) \cite{zirkel2015experience}, to collect the participants' self-reported engagement moods after each answer submission to HDU. The participants report their engagement moods through a Google form. After training each participant about the definition of each engagement mood, we ask them to identify their moods through one of the options shown in Table \ref{tab12356465}. The study lasts for a period of two weeks long, from September 18th to October 2nd, 2020, and 63 responses are recorded. Parallel to each self-reported engagement mood we receive, an HMM-based label is also generated through students' observed data from HDU. We notice that there is a 76.19\% agreement between the engagement mood labels extracted from the HMM and ESM. This is more than 50\% better than a random labeler with an accuracy of 20\%.

Interestingly, none of the participants mention any engagement moods other than the five engagement moods we have extracted. However, we predict that there would be more personalized and detailed engagement moods with an increased number of participants \cite{saenz2011community}.

\begin{table}[t!]
\small
  \centering
    \begin{tabular}{lc}
    \toprule
    Engagement Mood Options: & Responses \\
    \hline
    (1) I am looking for a challenging question. (Challenge-seeker) & 46.03\% \\
    (2) I am looking for a specific question. (Subject-seeker) & 19.04\% \\
    (3) I am looking for an interesting (non-challenging) question. (Interest-seeker) & 12.69\% \\
    (4) I want to use the platform for purposes other than learning. (Joy-seeker) & 11.11\% \\
    (5) I want to solve a random question. (Non-seeker) & 11.11\% \\
    (6) None of the above (explain) & N/A \\
  \bottomrule
    \end{tabular}%
\caption{Engagement mood options participants could choose from.}\label{tab12356465}%
\end{table}%
\section{Engagement Moods and Student Dropouts (RQ2)}
\begin{figure}
     \centering
     \begin{subfigure}[b]{0.45\textwidth}
         \centering
         \includegraphics[width=\textwidth]{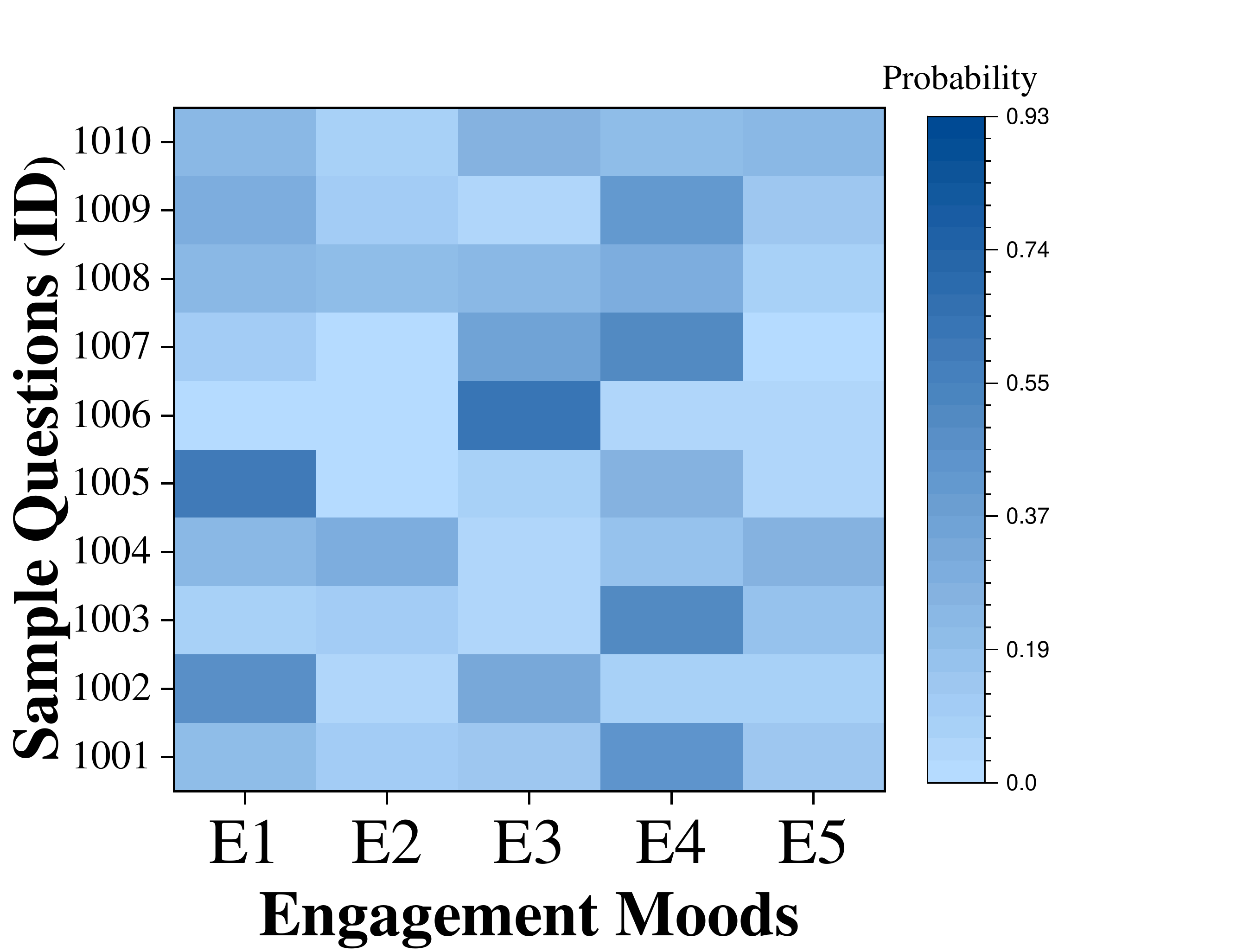}
         \caption{}~\label{fig:XXX}
     \end{subfigure}\hspace{0.01\textwidth}
          \begin{subfigure}[b]{0.45\textwidth}
         \centering
         \includegraphics[width=\textwidth]{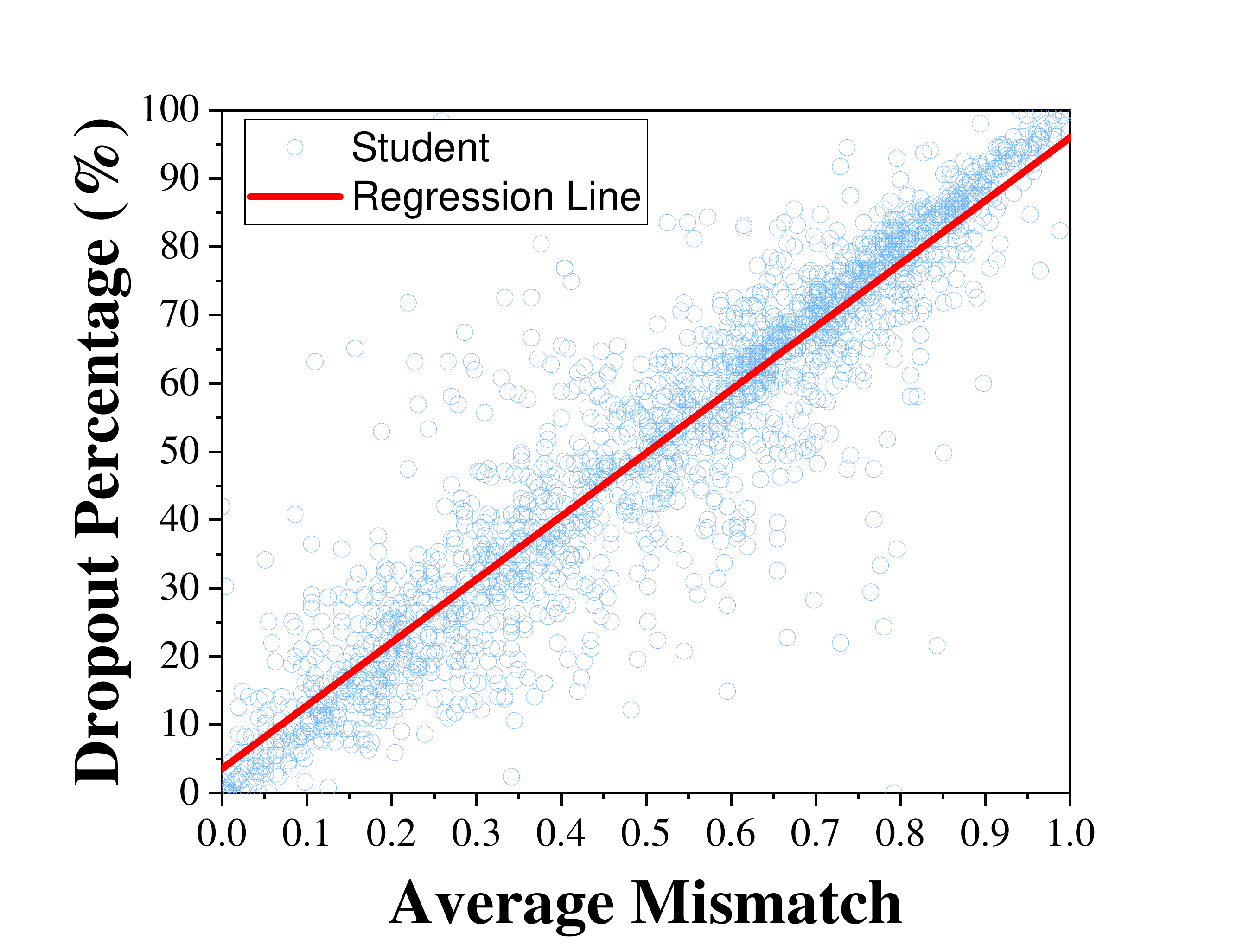}
         \caption{}~\label{fig:XXX2}
     \end{subfigure}
        \caption{(a) Finding the question associativity for sample questions, (b) The linear correlation between the average question mismatch and the percentage of student dropouts from the platform}~\label{fig:allTESTH}
\end{figure}

Next, we use the optimal \scalebox{0.90}{$\lambda_{\scalebox{0.50}{HMM}}$} to find the most probable engagement mood sequence \scalebox{0.90}{$X_{z}=\{x\in Z\}$} for each student with respect to their observed sequence \scalebox{0.90}{$O$} so as to maximize the \scalebox{0.90}{$Pr(X_{z}|O, \lambda_{\scalebox{0.50}{HMM}})$} (inferences at \cite{forney2005viterbi}). Furthermore, we associate every question \scalebox{0.90}{$q_{j}$} on the platform a distribution \scalebox{0.90}{$Q_{j}$} based on its probabilities for receiving answers when students are in different hidden states. Here, the index \scalebox{0.90}{$j$} refers to the identity number of the questions on the platform. We define the average question mismatch as the probability that a question does not match with the current engagement mood of a student. Illustratively, the question mismatch is the complement of the question's associativity measure, as is represented in Figure \ref{fig:XXX}.

In this subsection, we test the null hypothesis that the average question mismatch has no correlation with the percentage of student dropouts. The regression analysis shown in Figure \ref{fig:XXX2} reveals the positive coefficient factor of $92.45$, and the Pearson's r value of $0.927$ with the significance of $P_{value}<0.01$ between the average question mismatch and the student dropouts. Therefore, the null hypothesis is rejected, and the alternative hypothesis is accepted. That is to say, as the average question mismatch increases, the percentage of students who drop out also increases. Furthermore, it can be inferred that the students in each engagement mood have a collective preference for answering questions if from which they deviate, their risk of dropping out also increases.
\section{Dropout Prediction with Engagement Moods (RQ3)}
We suggest using engagement moods extracted from the HMM with an LSTM network to predict student dropouts in QPs more precisely. We refer to this hybrid machine learning architecture as Dropout-Plus (DP). 

We use \textit{Keras} in Python to build an attention-based Long Short-Term Memory (LSTM) network to detect the student dropouts \cite{sato2019early}. The features we input into the network at every time \scalebox{0.90}{$t$} the student submits an answer to the QP are of two types: 1) We call the set of non-handcrafted features the common features that are directly acquired from student behavior. The common features are often shared among different QP platforms and include the features of student's membership period, rank, nationality, acceptance rate, error type distributions, and the average time gap between submissions; and 2) the set of preprocessed features the HMM renders. In order to feed data into the network more effectively and reduce the effect of correlated features, we use a fully connected feed-forward neural network to combine the features and get a distributional feature set to train our model \cite{Yang:2018:IKY:3219819.3219821}.

We compare the performance of DP for predicting student dropouts with five competitive baselines picked from the literature. Since there are no preceding dropout prediction models for QPs before our work, we pick our baselines from previous MOOC and CQA studies \cite{Nagrecha:2017:MDP:3041021.3054162, Pudipeddi:2014:UCF:2567948.2576965}. To keep the comparisons relevant and unbiased, we avoid the models where it is not clear how to match MOOC or CQA features to QPs. The baselines include:

\noindent$\bullet$ \textbf{XGBoost}: Extreme Gradient Boosting (XGBoost) algorithm is one of the most dominant machine learning tools for classification and regression \cite{miri2020evaluating, kwak2019voice, wang2019designing, murauer2018detecting}. It comprises a collection of base decision tree models that are built sequentially, and their final results are summed together to reduce the bias \cite{chen2016xgboost}. Each decision tree boosts attributes that led to misclassification of the previous decision tree. 

\noindent$\bullet$ \textbf{Random Forest}: Random Forest is another decision tree-based ensemble algorithm that has a rich literature in HCI and CSCW communities \cite{rezapour2017classification, maity2017detection, kim2019will, shin2020guessing}. However, different from XGBoost, Random Forest combines the decision trees all uniformly by using an ensemble algorithm known as bootstrap aggregation \cite{bishop2006machine, breiman2001random}. In other words, every decision tree is independent of the others, and thus the final classification result is resolved based on a majority voting \cite{bishop2006machine}.
    
\noindent$\bullet$ \textbf{Decision Tree (DT)}: As a base model of Random Forest and XGBoost algorithms, DT's performance sometimes exceeds the two \cite{Nagrecha:2017:MDP:3041021.3054162}. However, even a small change in data would dramatically reshape the model, which is an adverse point. Nevertheless, we add DTs in our analysis to have a more comprehensive set of baselines \cite{Dror:2012:CPN:2187980.2188207, Nagrecha:2017:MDP:3041021.3054162, Pudipeddi:2014:UCF:2567948.2576965}.
    
\noindent$\bullet$ \textbf{Logistic Regression}: As a popular statistical tool in HCI and CSCW quantitative analysis \cite{masaki2020exploring, che2018fake, kim2019will, fortin2019detecting}, Logistic Regression in its basic form uses a logistic (sigmoid function) function to model a binary dependent variable \cite{masaki2020exploring}. In our work, the dependent variable is the dropout's outcome \cite{masaki2020exploring}.

\noindent$\bullet$ \textbf{SVM (RBF Kernel)}: Support Vector Machine (SVM) is another supervised algorithm that can be applied for both classification and regression purposes \cite{bishop2006machine}. SVM tries to identify hyperplanes (boundaries) that can separate all data points into groups with high margins \cite{benerradi2019exploring}. Gaussian Radial Basis Function (RBF) is one of the most common kernel functions researchers apply to train their models \cite{wang2019atmseer, xuan2016lbsnshield, liu2020data}. 
    
We also run an ablation study (see \cite{joseph2017girls, gardiner1974psychology}) with each baseline by including or excluding Engagement Features (EF) produced by the HMM. More precisely, EF includes students' engagement moods and their questions' associativity features after making each submission during the observation period. However, all of the models apply the common features we introduced before for model training. We remind the reader that the common features include student’s membership period, rank, nationality, acceptance rate, error type distributions, and the average time gap between submissions. All of the baselines are implemented through  \textit{sklearn} module in Python.  

Table \ref{tab:AblResTab1} summarizes the results of our analyses based on 10-fold cross-validation \cite{bishop2006machine, rodriguez2009sensitivity}. The results of our analyses show that DP outperforms all the other baselines we have suggested. Moreover, the models with HMM Engagement Features (EF) have all performed better than models without EF. In fact, adding EF has leveled up the performance of all models close to DP's performance, which can attest that the information about engagement moods can improve dropout prediction. 

\begin{table}[t!]
\begin{center}
  \includegraphics[width=0.95\linewidth]{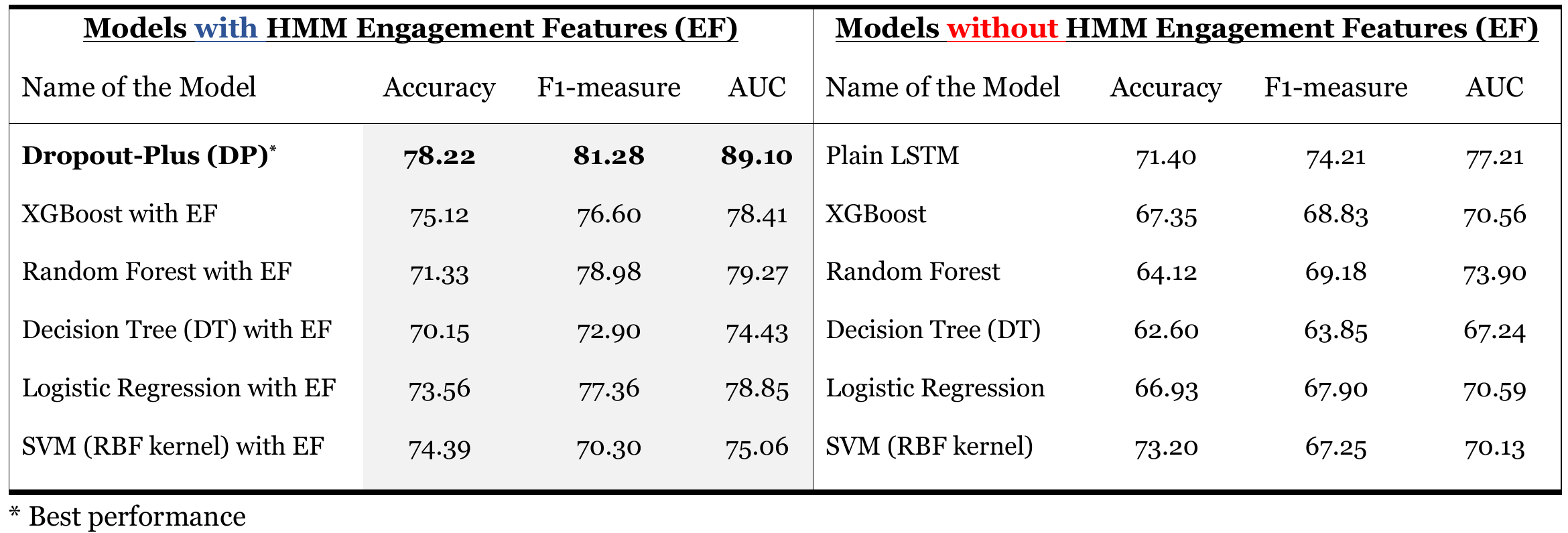}\\
\end{center}
\caption{Performances of different methods under different
evaluation metrics (all reported in  \%).}~\label{tab:AblResTab1}
\end{table}
\section{Discussion}\label{DDDD4}
To the best of our knowledge, this research is the first work that characterizes the students' engagement moods and sets the dropout prediction baseline for the QP platforms. Rendering and exposing student engagement moods is the sweet spot of our work, and the implications can provide practical insights for online learning professionals to manage students' behaviors better and tailor their services accordingly. 

\subsection{Engagement Mood Dynamics}
One of the difficulties of studying human behavior in HCI and CSCW studies is that people change their behaviors dynamically over time \cite{dos2018does, sleeper2015would, alghamdi2019crowd, lee2019commitment}. Therefore, studying the dynamics between student engagement moods is an essential part of understanding students' behavior.

Based on the observations from students' engagement mood changes in our dataset, we have calculated the transition probabilities between different engagement moods, as shown in Table \ref{tab:transitF}. Therefore, these probabilities show a frequentist perspective of engagement mood transitions for a restricted time period (i.e., 172 days) \cite{spanos2019probability}. As is shown, the engagement moods are well-distinguished with respect to their probabilistic distributions. The general dynamics from Table \ref{tab:transitF} show that holistically students have the highest probability of getting into the interest-seeker mood and the least probability for getting into the challenge-seeker mood.

Interestingly, the students in the challenge-seeker mood do not become joy-seekers or non-seekers. We also realize that the joy-seeker students do not become challenge-seekers or subject-seekers. They look for easy questions, and the subjects seem not to interest them. 
Furthermore, we observe that the challenge-seeker students have the highest probability of becoming interest-seekers. On the other hand, interest-seeker students do not become challenge-seekers at any time. However, this observation may be the result of the inappropriate question assignments to the students. According to our findings, subject-seeker students are also not terribly motivated to change their moods, which sounds reasonable because of the directed and specific-context nature of the questions they choose to answer. Interest-seekers are more likely to become non-seekers and vice-versa. Since finding the programming topics of interest is often the main intention for most of the student explorations in QPs, a fit question recommender system can be advantageous for students' satisfaction and quality of experience \cite{mogavi2019hrcr,Xia:2019:PPI:3290605.3300864}. Finally, among all the resolved engagement moods, students in the joy-seeker mood have the least tendency to change their moods. This implies that the platform resources are largely at risk of being wasted unless we find and guide the joy-seeker students on time \cite{Wasik:2018:SOJ:3177787.3143560}.

\begin{table}[t!]
\centering
\includegraphics[scale=0.6]{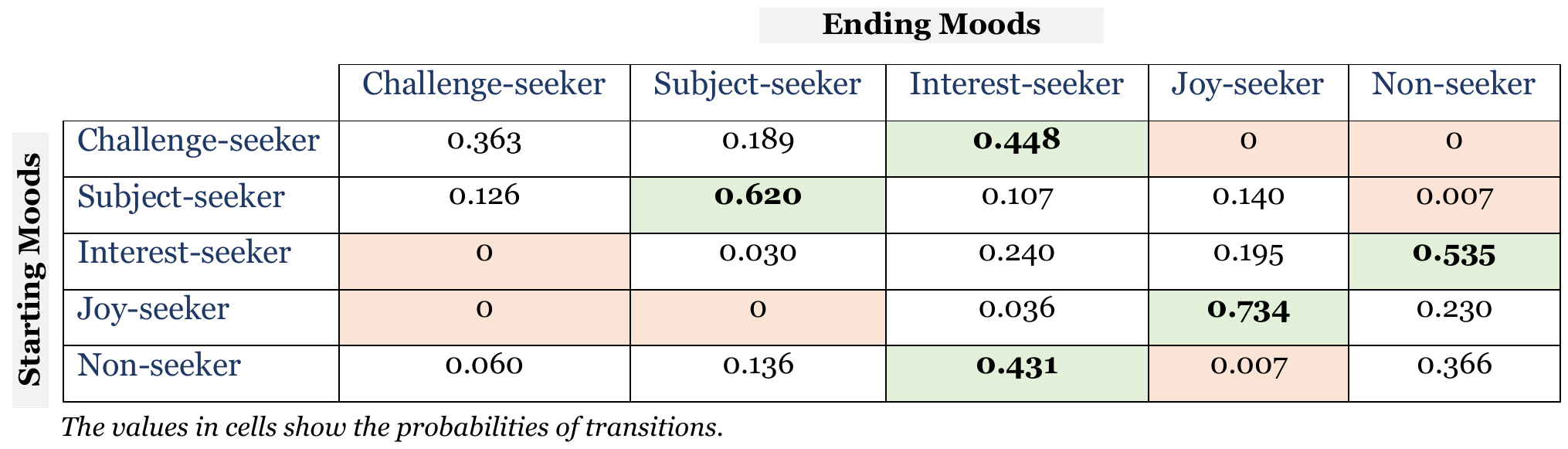}
\caption{The transition matrix of moving between different engagement moods}~\label{tab:transitF}
\end{table}

\subsection{Dominant Engagement Moods}
We define a dominant engagement mood as a mood that has happened most frequently in a student's engagement mood trajectory. As is shown in Table \ref{fig:dominant1}, the majority of the students ($30.23\%$) are often in the non-seeker mood. They also hold the highest average mismatch measure ($78.21\%$) for the questions they answer, and as is expected, they hold the worst dropout rate ($81.03\%$) among all of the other engagement moods. According to our findings, the average mismatch measure ($23.05\%$) and dropout rate ($13.64\%$) are the least for the dominantly challenge-seeker students. Also, although we saw that the students in the joy-seeker mood have the least tendency to change their moods, the students who are dominantly in this mood have the second highest dropout rate ($73.4\%$) on the platform. It is likely that they get bored or encounter some technical issues that force them to stop gaming the platform.

\begin{table}[t!]
\small
\begin{center}
  \includegraphics[width=0.85\linewidth]{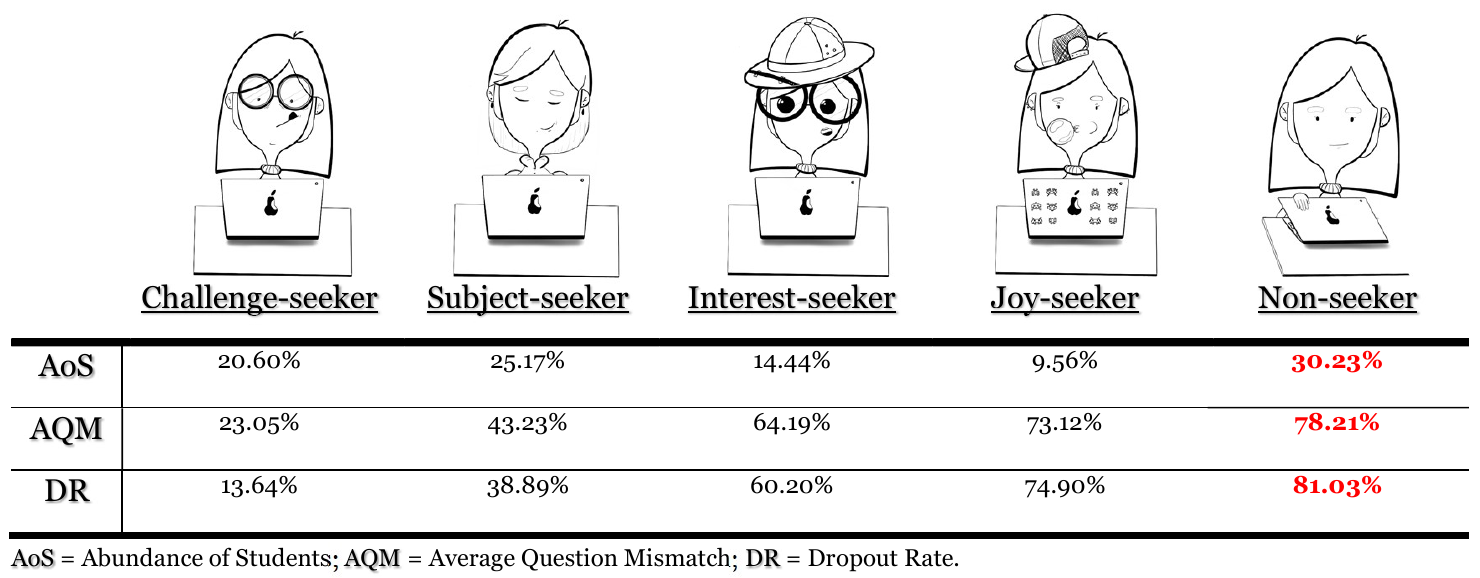}\\
\end{center}
\caption{Dropout analysis according to the dominant engagement moods of the students}~\label{fig:dominant1}
\end{table}

\subsection{Design Suggestions}\label{SSSS4}
Based on our findings, we provide some design suggestions for online learning professionals to better control student dropouts in QPs. 

\noindent$\bullet$ \textbf{Recommender systems.} Current recommender systems do not consider \textit{when} is the right time for a student to learn something, they just consider what students want to learn \cite{Xia:2019:PPI:3290605.3300864}. The problem holds for many educators as well who do not know when it is the right time to engage students for studying a specific topic. Our resolved engagement moods can play a role as an assistant to address this issue better. For example, if a topic is difficult and challenging, we would recommend asking students to solve it when they are in a challenge-seeker mood. If students are in a joy-seeker mood, we would suggest educators to skip asking questions in that session because it might actually result in adverse and unwanted educational outcomes in students, such as feeling frustrated or disliking the subject matter of that course \cite{baker2008students}.
Combining learning path recommenders or personal teaching (coaching) styles with engagement mood exposers would probably enhance the quality of education and deter dropouts as students would feel more satisfied with what they learn and do \cite{suhre2007impact}. 

\noindent$\bullet$ \textbf{Strategic dropout management.} Before our work, researchers have not paid much attention to \textit{who} is dropping out \cite{10.1007/978-3-030-23207-8_13, whitehill2017mooc}, but we think that this is an important question to be answered for managing student dropouts more strategically. In contrast with the existing practice in dropout studies, we believe not all of the dropouts are equally bad. For example, imagine an educator who is undecided about which types of interventions she should apply to retain students from dropping out. She is confused about customizing the system either according to challenge-seekers' or joy-seekers' tastes. We would recommend prioritizing challenge-seekers as joy-seekers would most probably game the system without learning anything. Joy-seekers might also mischievously have adverse side effects on other students' experiences and make them feel disappointed \cite{cheng2017anyone, baker2008students}. For more complex dropout scenarios, educators can also heed to the abundance of students in each group and the transition probabilities between different engagement moods to make better decisions.

\noindent$\bullet$ \textbf{Gamification and reinforcement tools.} According to Skinner's refined Law of Effect, reinforced behaviors tend to repeat more in the future, and those behaviors which are not reinforced tend to dissipate over time \cite{bhutto2011effects}. Therefore, educators can use gamification incentives such as badges, points, and leaderboards to steer students' behaviors \cite{anderson2013steering}. We suggest QP designers to consider student engagement moods for designing their gamification mechanics \cite{tondello2019gameful}. For example, if educators are preparing students for difficult exams like ACM programming contests, they can reinforce challenge-seekers by providing bigger gamification prizes \cite{zhang2016social}. Likewise, negative reinforcement can be applied to diminish non-productive behaviors like being in a joy-seeker mood. Besides that, QP designers can provision special gamification mechanics or side games to satisfy students' playful intentions somewhere else.

\noindent$\bullet$ \textbf{Putting students into groups.} At this point, collaboration and social communication are hugely missing in the context of QPs, and instead, all the website functions only promote competency among students. Although competency can provide an initial motive to attract students \cite{bartel2015towards}, many educators believe that collaboration and social communication among students are also needed to better achieve the intended educational outcomes \cite{janssen2020applying, tucker2016collaboration}. We suggest QPs to add affordances to support student communication and social collaboration as well. Meanwhile, student engagement moods can help to find initial common attitudes among students and form groups. For example, if one student is a subject-seeker, there is a good chance that this student would successfully fit into one group with other subject-seeker students. This rule is known as Homophily in social networks \cite{mcpherson2001birds}. Nevertheless, more in-depth studies are required to figure out which combination of student engagement moods work and match better together.

\subsection{Limitations}~\label{AReflection1}


As with any study, there are several limitations and challenges in this study. First, the platform data we analyzed is cross-sectional and is restricted in its size and time, but not type. We find this problem to be a commonplace issue in many related studies as well \cite{Nagrecha:2017:MDP:3041021.3054162,mogavi2019hrcr}. Also, we want to emphasize the difficulty of working with student data and the scarceness of datasets about QP platforms, which are relatively new research subjects in the context of educational data mining. Regarding the model limitations, although HMM can successfully profile the student behavior in a few hidden states, it is often a time-consuming task to characterize the resolved hidden states. For example, in this research, we have spent more than twelve hours to carefully visualize and compare the probabilistic distributions behind the hidden states considering different aspects and features to finally announce our engagement mood typology. 
We should also explain that since hidden states are found by an estimation procedure, different platforms might result in different numbers of hidden states. Generally, we expect these states to be semantically close to what we have introduced in this work. We emphasize that this feature should be viewed more as a positive point for future work, which leaves more complex engagement moods to be mined and compared with our extracted typology. Hence, it remains as future work to cross-validate our work on other QP platforms such as Timus, quera.ir, CodeChef, and the like. Moreover, it would be a particularly interesting direction to examine the effect of the HMM's state space granularity on the performance of the dropout predictions. 

\section{Conclusion}\label{Conclusion}
We used the powerful tool Hidden Markov Models (HMMs) to expose  underlying student engagement moods in QP platforms and point out that the mismatch between students' engagement moods and the question types they answer over time can significantly increase the dropout risk. Furthermore, we developed a novel and more accurate computational framework called Dropout-Plus (DP) to \textit{predict} student dropouts and \textit{explain} the possible reasons why dropouts happen in QP platforms. However, we believe there is still a long path in front of HCI and CSCW researchers to fully understand dropouts on different educational platforms. Our future work includes developing a more exact time prediction for student dropouts and enriching the explanations to the question of ``why dropouts happen?'' Finally, this study can benefit researchers and practitioners of online education platforms to promote their work by understanding student dropouts more profoundly, building better prediction models, and providing more customized services.
\begin{acks}
This research has been supported in part by project 16214817 from the Research Grants Council of Hong Kong, and the 5GEAR and FIT projects from Academy of Finland. It is also partially supported by the Research Grants Council of the Hong Kong Special Administrative Region, China under General Research Fund (GRF) Grant No. 16204420.
\end{acks}

\bibliographystyle{ACM-Reference-Format}
\bibliography{sample-base}









\end{document}